\documentclass[journal]{IEEEtran}

\IEEEoverridecommandlockouts
\makeatletter
\def\ps@headings{%
	\def\@oddhead{\mbox{}\scriptsize\rightmark \hfil \thepage}%
	\def\@evenhead{\scriptsize\thepage \hfil \leftmark\mbox{}}%
	\def\@oddfoot{}%
	\def\@evenfoot{}}
\makeatother
\usepackage{subfigure}
\usepackage{colortbl}
\usepackage{bm}

\usepackage{graphicx}  
\usepackage{url}       

\usepackage{amsmath}   
\usepackage{cite}

\usepackage{amsfonts,amssymb}
\usepackage{cite}
\usepackage{stfloats}

\usepackage{cases}
\usepackage{algorithm}
\usepackage{multirow}
\usepackage{algorithmic}
\usepackage{stfloats}
\usepackage{epstopdf}

\makeatletter

\newcommand{\Rmnum}[1]{\expandafter\@slowromancap\romannumeral #1@}
\makeatother

\UseRawInputEncoding

\newcommand{\ls}[1]
{\dimen0=\fontdimen6\the\font
	\lineskip=#1\dimen0
	\advance\lineskip.5\fontdimen5\the\font
	\advance\lineskip-\dimen0
	\lineskiplimit=.9\lineskip
	\baselineskip=\lineskip
	\advance\baselineskip\dimen0
	\normallineskip\lineskip
	\normallineskiplimit\lineskiplimit
	\normalbaselineskip\baselineskip
	\ignorespaces
}

\begin{document}
	\title{Reconfigurable Intelligent Surface Equipped UAV in Emergency Wireless Communications: A New Fading-Shadowing Model and Performance Analysis}
	\vspace{10pt}
\author{	\IEEEauthorblockN{Yinong Chen, \emph{Student Member}, \emph{IEEE}, Wenchi Cheng, \emph{Senior Member}, \emph{IEEE}, and Wei Zhang, \emph{Fellow}, \emph{IEEE}}\vspace{-17pt}
		\thanks{Part of this work was presented in IEEE International Conference on Communications, 2022\cite{myf}.
		
		Yinong Chen and Wenchi Cheng are with the State Key Laboratory of Integrated Services Networks, Xidian University, Xi'an, 710071, China (e-mails: yinongchen@stu.xidian.edu.cn, wccheng@xidian.edu.cn). 
		
		Wei Zhang is with the School of Electrical Engineering and Telecommunications, University of New South Wales, Sydney, NSW 2052, Australia (e-mail:
		w.zhang@unsw.edu.au).}
	}
		
		

	%

	\maketitle

	\begin{abstract}
	Communication infrastructure is often severely disrupted in post-disaster areas, which interrupts communications and impedes rescue. 
	Recently, the technology of reconfigurable intelligent surface (RIS)-equipped-UAV has been investigated as a feasible approach to assist communication under such conditions. 
	However, the channel characteristics in the post-disaster area rapidly change due to the topographical changes caused by secondary disasters and the high mobility of UAVs.  
In this paper we develop a 
new fading-shadowing model to fit the path loss caused by the debris. 
	Following this, we derive the exact distribution of the new channel statistics for a small number of RIS elements and the approximate distribution for a large number of RIS elements, respectively. 
	Then, we derive the closed-form expressions for performance analysis, including average capacity (AC), energy efficiency (EE), and outage probability (OP). 
    Based on the above analytical derivations, we maximize the energy efficiency by optimizing the number of RIS elements and the coverage area by optimizing the altitude of the RIS-equipped UAV, respectively. Finally, simulation results validate the accuracy of derived expressions and show insights related to the optimal number of RIS elements and the optimal UAV altitude for emergency wireless communication (EWC).

	\end{abstract}

\section{Introduction}
\IEEEPARstart{A}{fter} natural disasters such as earthquakes, tsunamis, and mountain torrents, the terrestrial communication infrastructure is often destroyed in the post-disaster areas, resulting in the disruption of communication and 
severely impeding rescue operations. 
With the development of the unmanned aerial vehicle (UAV) technology, UAV-assisted wireless communication, as a promising scheme for future wireless networks, attracts increasing attentions\cite{EWC,EWCUAV,Harsh}. 
Due to their high mobility, easy operation, and affordable price, UAVs can be rapidly deployed as base stations (BSs) or relays in post-disaster areas, supporting the temporary communication networks\cite{Harsh,Yao,Cheng,3Ddeployment,UAVwang}. 
To date, a series of existing research work has focused on the statistical performance characterization and optimization for UAV-assisted wireless networks \cite{UAVrelay1,UAVrelay3,UAVopt}. 
The authors of \cite{UAVrelay1} 
analyzed the outage probability of a UAV-assisted relaying system under line-of-sight (LoS) and non line-of-sight (NLoS) connections with Rician and Rayleigh fading channels, respectively. 
For a UAV-assisted data-ferrying network, the authors of \cite{UAVrelay3} developed a mathematical framework to characterize the reliability, energy efficiency, and coverage probability with Rician fading aerial channels and optimized the data-ferrying distance to minimize both the outage probability and energy consumption.  
However, UAVs also have some typical shortcomings, such as limited resources and unclear channel model, which leads to low capacity and energy efficiency. 

To further increase the channel capacity and energy efficiency (EE) of wireless communication, reconfigurable intelligent surface (RIS), as an emerging technology, attracts much attention. RIS is composed of low-cost passive elements, which can intelligently control the propagation of impinging electromagnetic (EM) waves but do not amplify or introduce noise when reflecting signals, thus increasing signal coverage, maximizing signal-to-noise ratio (SNR), 
and optimizing beamforming\cite{RISmerit1,RISmerit2,RISmerit3,MIMO,ad1}. 
The RIS-aided system is not only proved to outperform relay-aided schemes in terms of both average capacity and outage probability in the cascaded channels\cite{ad2,ad5},  but also achieve diversity gains when RIS is employed as a transmitter\cite{ad3}.
Additionally, RIS can be installed on objects of any shape to meet different application scenarios\cite{RISmerit4}. 
 Recently, through the integration of an software-controlled-RIS on UAV, which is known as aerial RIS (ARIS), the wireless communication environment can be reconfigured, thereby reducing the deployment cost and considerably enhancing the performance of wireless communication network in terms of system throughput, coverage probability, energy efficiency, and spectrum efficiency\cite{ARIS1,ARIS2,ad4}. 
For future wireless communication, it is also required to develop a more flexible hardware architecture. Therefore, by fully reaping the aforementioned benefits, UAV-RIS-based wireless communication is expected to be a feasible scheme, especially under emergency wireless communication (EWC) scenario\cite{Yao,Harsh,RISzhang}. 

{Currently, the most commonly used RIS-based channels adopted LoS/NLoS connections or Rician, Rayleigh, and  Nakagami-$m$ fading \cite{UAVpath,RISrician,UAVRISNOMArician}. }
	However, the severe multipath fading and shadowing in post-disaster scenarios caused by unpredictable secondary disasters and complex topographical changes make it impossible for any of the aforementioned models to accurately capture both fading and shadowing simultaneously, thus rendering them unsuitable for EWC channels. 
Considering the impact of concurrent multipath fading and shadowing on wireless links, the authors of \cite{kappamufading,alphamufading} proposed several composite fading-shadowing models, which are mixtures of some classical multipath fading distributions with inverse gamma or inverse Nakagami-$\mathit{m}$ distributions. 
More recently, the authors of \cite{Fchannel,Fchannel3} proposed the Fisher-Snedecor $\mathcal{F}$ distribution to model both multipath fading and shadowing with Nakagami-$\mathit{m}$ and inverse Nakagami-$\mathit{m}$ fading, respectively. 
With lower complexity and a better fit to the practical data, this model can more effectively describe the characteristics of wireless channels with composite fading and shadowing. 
For MISO systems, the authors of \cite{Fomega} only considered the case of light shadowing for effective rate under the Fisher-Snedecor $\mathcal{F}$ fading channel. 
For wireless multiple access channels (MAC), the authors of \cite{Fcorrelated} utilized copula theory to derive the exact analytical expressions for the outage probability (OP) and average capacity (AC) in both correlated/independent Fisher-Snedecor $\mathcal{F}$ fading channels.  
For RIS-aided wireless communications, the authors of \cite{FIOT} derived closed-form expressions with regards to average bit error rate (BER) and OP for IoT networks, while the authors of \cite{Fsecurity,secureF} both considered a RIS-based secure wireless communication scenario. The authors of \cite{Fsecurity} evaluated security performance including average secrecy rate (ASC) and secrecy outage probability (SOP) of a RIS-aided MIMO system under two commmunication cases: communication without and with RIS. While the authors of \cite{secureF} proposed a cascaded Fisher-Snedecor $\mathcal{F}$ channel model and derived the closed-form and asymptotic expressions of SOP and ASC by exploiting the Laguerre expansion apprximation. 
{However, for conventional Fisher-Snedecor $\mathcal{F}$ model\cite{Fchannel,Fchannel3,Fomega,Fcorrelated,FIOT,Fsecurity,secureF}, when the channel experiences severe shadowing in the multi-antenna system (e.g., RIS-aided systems), there is a large error between the statistical result for the mean power and the mean power in the existing distribution equation. This results in the performance downgrading of AC, OP, and BER in harsh shadowing environments. 
In specific, \cite{Fcorrelated} only simulated the case when the fading and shadowing parameters  $m$ and $m_s$ simultaneous decrease, whereas \cite{Fomega,Fsecurity,secureF} ignored the change of $m_s$, making it challenging to evaluate the impact of $m_s$ on the simulation results. Also, when we reproduced the performance analysis from \cite{FIOT}, the simulation results revealed that the average BER increases as the shadowing becomes more serious when the mean power $\Omega$ is set to 1, which is in contrast to the simulation results presented in \cite{FIOT}.}

To solve the above-mentioned problem of accurately modeling both the severe multipath fading and shadowing under complex channel conditions, we propose a novel modified-Fisher-Snedecor $\mathcal{F}$ fading channel. Different from the conventional Fisher-Snedecor $\mathcal{F}$ fading channel, we address the issue of the large error between the statistical result of the mean power and the mean power in existing distribution equations under harsh shadowing conditions.  
Based on the composite channel model in the post-disaster scenario, we derive both the exact distribution of the new channel statistics with a small number of RIS elements and the approximate distribution with a large number of RIS elements to cover the full range of the number of RIS elements. 
	Then, the closed-form expressions are provided for the average capacity (AC), energy efficiency (EE), and outage probability (OP), including theoretical bounds on the expectation of SNR and OP. The derived results can provide insights into the fact that the performance is significantly improved by increasing the number of RIS elements under severe multipath fading and shadowing. 
In view of this, we first take the power consumption of each RIS element and the UAV carrying capacity constraints into account, aiming to maximize the EE by optimizing the number of RIS elements. Second, we formulate the coverage area optimization problem to maximize the service radius under the constraints of both the OP and the flight altitude of the RIS-equipped UAV. 
Numerical results show that the optimal number of RIS elements and altitude of the RIS-equipped-UAV scheme under our novel modified Fisher-Snedecor $\mathcal{F}$ fading channel can achieve a higher gain than the solutions obtained through the conventional Fisher-Snedecor $\mathcal{F}$ fading channel. 


The rest of this paper is organized as follows. Section~\ref{sec:System_model} describes the RIS-UAV-based system under the EWC scenario. 
Section~\ref{sec:Performance} presents the closed-form expressions of the new channel statistics 
and characterize the average capacity, energy efficiency, and outage probability. 
Section~\ref{sec:Optimization} formulates the optimization problems of maximizing energy efficiency for the number of RIS elements and maximizing the coverage area for the altitude of RIS-equipped-UAV. 
Section~\ref{sec:Numerical} numerically evaluates our optimal number of RIS elements scheme and altitude of RIS-equipped-UAV scheme for the novel modified-Fisher-Snedecor $\mathcal{F}$ channel as well as comparing with the conventional Fisher-Snedecor $\mathcal{F}$ channel. The paper concludes with Section~\ref{sec:Conclusion}. 

\section{System model}\label{sec:System_model}

\begin{figure*}[ht]
	\centering
	\includegraphics[scale=0.25]{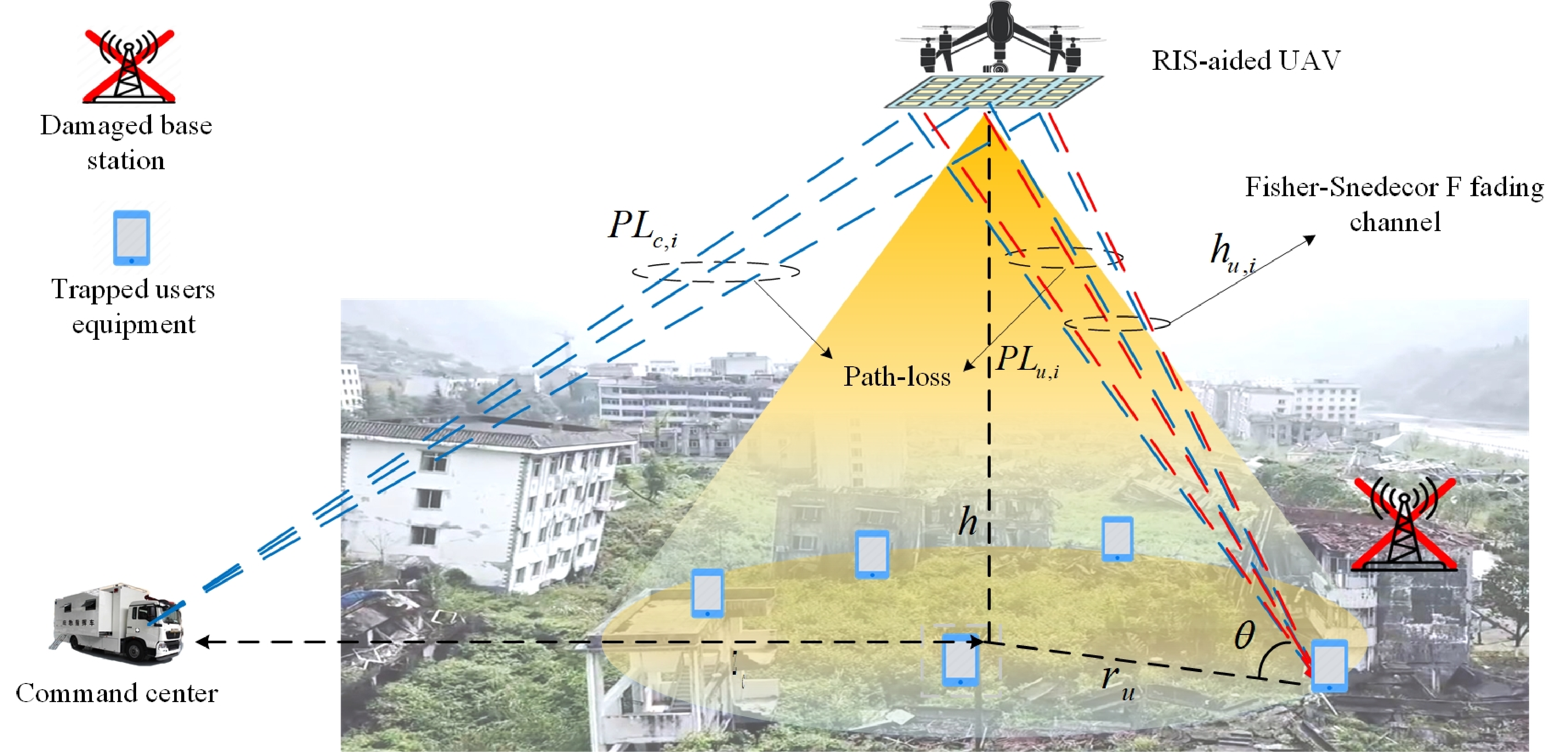}
	\vspace{-10pt}
	\caption{\centering{UAV-RIS-based system model under emergency wireless communication scenario.}}
	\vspace{-10pt}
	\label{Fig.model}
\end{figure*}

Figure~\ref{Fig.model} depicts an EWC system in the post-disaster area, in which the BSs are damaged and the trapped users are irregularly distributed. 
	As a result, an emergency communications vehicle that acts as the command center is set up to send/receive rescue instructions.
	Since the topography of the complex changes and ruins caused by the secondary disaster mostly blocks the direct links between the command center and the trapped user's equipment (UE), the UAV carrying a RIS with $N$ identical reflecting elements hovers over the post-disaster area to re-establish communications. As a result, the signal received by UE is a superposition of the reflected waves from RIS. \footnote{For the considered scenario, frequency division and time division are adopted as common and effective methods for the irregularly distributed trapped users in the post-disaster area. Hence, each trapped user can be considered separately as an optimization problem with its own assigned frequency band/time slot. 
	By reasonably choosing the location of the emergency communication vehicle, such as the highland with wide vision and no obstacles, the channel between the command center and the UAV can be regarded as mainly experiencing large scale fading, regardless of small scale fading. However, the severe multipath fading and shadowing caused by the ground ruins in the link between the UAV and UE can't be overcome in this way. 
} 
Without loss of generality, as illustrated in Fig.~\ref{Fig.model}, the received signal at the UE, denoted by $y$, can be expressed as follows:
\begin{equation}
y=\sqrt{P_s}\left(\sum_{i=1}^{N}\frac{h_{u,i}e^{-j\varphi_i}R_i}{\sqrt{PL_{c,i}PL_{u,i}}}\right)x+n,
\end{equation}
where $ P_s $ denotes the transmit power at the command center, $x$ denotes the normalized transmit signal with unit power, and $n$ represents the thermal noise, modeled as $\mathcal{C} \mathcal{N} \left(\mathbf{0}, N_0 \right)$. The term $R_i= A_i e^{\Phi_i}$ is the reflection coefficient of the $i$th reflecting element, where $A_i$ and $\Phi_i$ stand for the controllable amplitude and phase shift of each element, respectively, with $A_i=1$ for simplicity. The path loss between the command center and the $i$th element of the RIS and the path loss between the $i$th element of the RIS and the UE are denoted by $PL_{c,i}$ and $PL_{u,i}$, respectively.
The channel between the $i$th element of the RIS and UE is modeled as $h_{u,i} e^{-j\varphi_{i}}$, where $h_{u,i}$ and $\varphi_{i}$ represent the envelope and phase of the signal, respectively. 

\subsection{Path loss model}
For the link between the command center and the RIS-equipped UAV, the LoS communications mainly suffer from large-scale fading. Therefore, the path loss $PL_c$ \footnote{The distance between the command center and the $i$th RIS element is approximately equal to the distance between the command center and the UAV due to the small size of the UAV and RIS. The distance between the $i$th RIS element and UE is the same. In the following, $PL_{c,i}$ and $PL_{u,i}$ are both simplified as $PL_c$ and $PL_u$, respectively.} is modeled as 
\begin{equation}\label{plm1}
PL_c= d_c^\alpha, 
\end{equation}
with the path loss exponent $\alpha$ and the distance between the command center and RIS $d_{c}$. For the given flight altitude of the RIS-equipped-UAV $h$ and horizontal distance between the command center and RIS $z_c$, $d_{c}$ can be represented as follows:
\begin{equation}
d_{c}= \sqrt{z_c^2+h^2}.
\end{equation}
The link between the RIS-equipped-UAV and the UE adopts the new path loss model based on the statistical parameters of the underlying terrestrial environment\cite{LAP}. The path loss consists of two main propagation groups. The first group corresponds to receivers that favor a LoS condition. The second group corresponds to receivers that have NLoS but still receive signals via strong reflection and diffraction. 
Accordingly, the path loss, denoted by $PL_u^{{dB}}$ (measured in dB), is modeled as probabilistic LoS and NLoS links as follows:
\begin{equation}{\label{pl1}}
PL_u^{dB}=PL^{dB}_{_{LoS}}P_{_{LoS}}(\theta)+PL^{dB}_{_{NLoS}}P_{_{NLoS}}(\theta).
\end{equation}
The probability of LoS propagation, denoted by $P_{_{LoS}}(\theta)$, can be expressed as follows:
\begin{equation}
P_{_{LoS}}(\theta)=\frac{1}{1+ae^{-b(\theta-a)}},
\end{equation}
where $\theta$ is the elevation angle. The parameters $a$ and $b$ called the S-curve parameters, are specific to the environment being simulated in Tables I-II\cite{LAP}. Correspondingly, the probability of NLoS propagation can be expressed as $P_{_{NLoS}}(\theta)=1-P_{_{LoS}}(\theta)$. 
The mean path loss of the two propagation groups is composed of free space path loss (FSPL) and excessive path loss, which can be modeled as 
\begin{equation}
PL^{dB}_{\chi}=20\text{lg}\frac{4\pi d_u}{\lambda}+\eta_{_{\chi}}, \chi \in \left\{LoS, NLoS\right\},
\end{equation}
where $\lambda$ is the wavelength. The term $d_{u}$ is the distance between the RIS-equipped UAV and trapped users, which can be represented as $d_{u}={h}/{\text{sin}\theta=\sqrt{h^2+r_u^2}}$ with $r_u$ being the largest 2D distance between the RIS-equipped-UAV and a UE. 
The term $\eta_{_{\chi}}$ represents excessive path loss, which largely depends on the propagation group rather than the elevation angle. 
Hence, Eq.~(\ref{pl1}) can be rewritten as
\begin{equation}\label{plmodel}
PL_u^{dB}(\theta,h)=20\text{lg}\frac{4\pi h }{\lambda{\text{sin}\theta}}+\frac{(\eta_{_{LoS}}-\eta_{_{NLoS}})}{1+ae^{-b(\theta-a)}}+\eta_{_{NLoS}}.
\end{equation}

\subsection{Multipath-Shadowing Composite Fading Model}
For the considered EWC system, we amend the channel model in \cite{Fchannel} to accurately describe both the serious multipath fading and shadowing. The envelope of the received signal $h_{u,i}$ is defined as $h_{u,i}=XY$, where $X$ represents the RMS (Root Mean Square) power of the received signal affected by shadowing. $Y=\sum_{i=1}^{M}{(p_i^2+q_i^2)}$ is the multipath coefficient, where $M$ is the number of clusters of multipath. The terms $p_i$ and $q_i$ denote the in-phase and quadrature components of the multipath waves of cluster $i$, which are mutually independent Gaussian RVs with the expectation satisfying $\mathbb{E}[p_i]=\mathbb{E}[q_i]=0$ and  $\mathbb{E}[p_i^2]=\mathbb{E}[q_i^2]=\sigma^2$. $X$ follows the inverse Nakagami-$\mathit{m}$ distribution\cite{Fchannel}, the PDF of which is given by 

\begin{equation}
 f_X(x)=\frac{2m_s^{m_s}}{\Gamma(m_s){\Omega_s}^{m_s}x^{2m_s+1}}\text{exp}\left(-\frac{m_s}{{\Omega_s}x^2}\right),
\end{equation}
where $m_s$ represents the parameter of shadowing and $\Omega_s=\mathbb{E}[x^2]$ denotes the the RMS power affected by shadowing. 
$Y$ follows the Nakagami-$\mathit{m}$ distribution, the PDF of which is given by
\begin{equation}\label{fy}
f_Y(y)=\frac{2m^{m}y^{2m-1}}{\Gamma(m){\Omega_m}^{m}}\text{exp}\left(-\frac{my^2}{\Omega_m}\right),
\end{equation}
where $m$ represents the parameter of multipath fading and $\Omega_m=\mathbb{E}[y^2]=2M\sigma^2$ denotes the the mean power of the multipath component.  
Consequently, applying $f(h_{u,i})=\int_{0}^{\infty}\frac{1}{|X|}f_X(x)f_Y(\frac{h_{u,i}}{x})dx$ into Eq.~(\ref{fy}), the PDF of the composite signal envelope can be derived as follows:
	\begin{equation}\label{hPDF}
	f(h_{u,i})=\frac{2(m\Omega_s)^m(m_s\Omega_m)^{m_s}{h_i}^{2m-1}}{B(m,m_s)[m\Omega_s{h_i}^2+m_s\Omega_m]^{m+m_s}},
	\end{equation}
where $B(a, b) $ is the
beta function with $B(a,b) =\frac{\Gamma(a)\Gamma(b)}{\Gamma(a+b)}$\cite
[Eq.~(8.384.1)]{Integrals} and $\Gamma(b) = \int_{0}^{\infty}{t^{b-1}e^{-t}}dt
$ is the gamma function \cite[ Eq.~(8.310)]{Integrals}
. 
Let $t={h_i^{2}\Omega_s}/{\Omega_m}$, $t$ follows a $\mathcal{F}$ distribution with parameters $2m$ and $2m_s$ after performing variable changes and algebraic calculations. As $m, m_s \rightarrow 1$, the channel undergoes heavy fading and shadowing. When $m_s \rightarrow \infty$, the shadowing eventually disappears, indicating that the Fisher-Snedecor $ \mathcal{F} $ fading reduces to Nakagami-$m$ fading\cite{Fchannel,Fchannel3}. The mean power of the composite signal envelope can be expressed as $\Omega=\Omega_m\Omega_s$ in this case. \footnote{As described in previous literature\cite{Fchannel}, $\Omega_s$ is set equal to unity, i.e., $\Omega_s=1$. Therefore, the mean power of the signal can be expressed as  $\Omega=\Omega_m$, which is only related to the mean power of the multipath component. 
However, by using \cite[Eq.~ (3.194.3)]{2007Table}, the statistical result of mean power can also be calculated as $\mathbb{E}[h_{u,i}^2]=\frac{m_s}{m_s-1}\Omega$. Apparently, as the channel undergoes heavier shadowing, i.e. $m_s\to 1$, the mean power will be increasingly larger than $\Omega$. This contradicts the actual channel condition, leading to performance downgrading in harsh shadowing environments. 
 As a result, we propose a novel modified-Fisher-Snedecor $\mathcal{F}$ fading channel model with the definition of $\Omega$ as $\Omega=\Omega_m\Omega_s$, which indicates that the mean power is affected by both the multipath and shadowing components.}


The RIS controller is assumed to intelligently adjust the phase shift of each element to satisfy the equation $\Phi_i=\varphi_i$ through phase matching, thus achieving the maximum received SNR $\gamma$.  { In this case, the channel state information (CSI) of all channels involved is perfectly known at the receiver through various channel  estimation methods\cite{estimationfinal}.} In addition, the quasi-static flat-fading model is assumed for all channels. Subsequently, $\gamma$ can be expressed as follows:


\begin{equation}\label{SNR}
\gamma=\frac{P_s}{N_0}\left[\sum_{i=1}^{N}\frac{{ \left|{h_{_{u,i}}}\right|}}{\sqrt{ PL_{\tiny{C,i}}PL_{\tiny{U,i}}}}\right]^2=\frac{\gamma_0}{PL_{\tiny{C}}PL_{\tiny{U}}}A^2,
	\end{equation}
where $\gamma_0={P_s}/{N_0}$ is the average transmit SNR and $A=\sum_{i=1}^{N}\left|{h_{_{u,i}}}\right|$ represents the envelope of the received signal from all the reflecting elements. Here, $A$ is a sum of $N$ i.i.d. RVs.

\begin{figure}[t] 
	\vspace{-15pt}
	\centering 
	\includegraphics[scale=0.5]{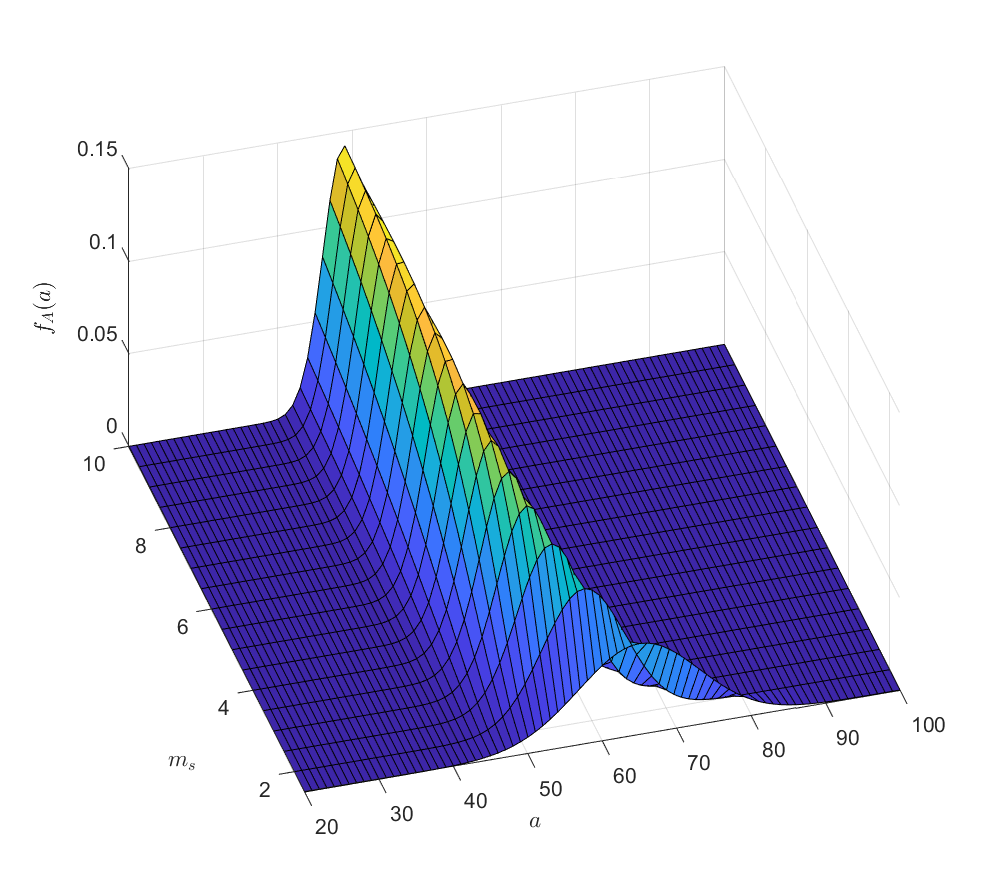}
	\vspace{-15pt}
	\caption{The PDF of the conventional $\mathcal{F}$ fading channel versus the channel characteristics of envelope of the received signal $A$ and shadowing parameter $m_s$.} 
	\vspace{-15pt}
	\label{Fig.conventionalsurf} 
\end{figure}

\begin{figure}[t] 
	\centering 
	\includegraphics[scale=0.42]{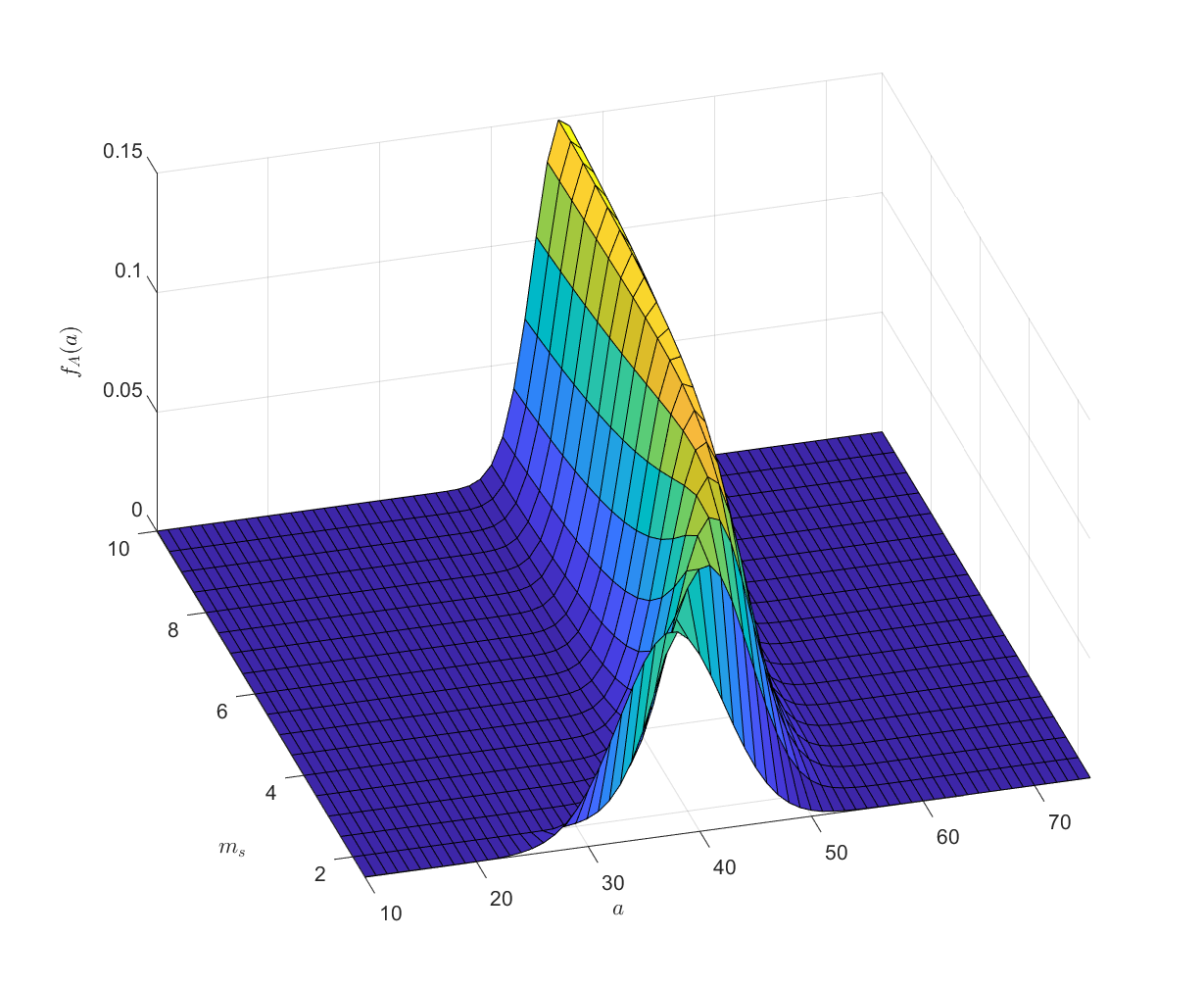}
	\vspace{-15pt}
	\caption{The PDF of the modified-$\mathcal{F}$ fading channel versus the channel characteristics of envelope of the received signal $A$ and shadowing parameter $m_s$.} 
	\vspace{-15pt}
	\label{Fig.newsurf} 
\end{figure}

{ To validate the accuracy of our proposed modified-$\mathcal{F}$ fading channel model, we depict the PDF of the conventional $\mathcal{F}$ fading channel and modified-$\mathcal{F}$ fading channel versus the channel characteristics of signal envelope $A$ and shadowing parameter $m_s$ in Figs.~\ref{Fig.conventionalsurf} and \ref{Fig.newsurf}, respectively. As shown in Fig.~\ref{Fig.conventionalsurf}, the range of $A$ is gradually concentrated to the left as $m_s$ increases, which means the signal envelope tends to decrease as the channel undergoes more slight shadowing. Clearly, it is in contrast to the actual channel conditions. However, it can be seen from Fig.~\ref{Fig.newsurf} that when $m_s$ increases, the PDF of $A$ shifts to the right, which is consistent with the actual channel situation.} 
\subsection{Power consumption model}
The total power consumption of the EWC system 
is expressed as\cite{RISCONSUMPTION}
\begin{equation}
P_{tot}=\mu P_s+p_c+p_h+p_{_{RIS}},
\end{equation}
 where $\mu=\upsilon^{-1}$ with $\upsilon$ being the efficiency of the transmit power amplifier, $p_c$ represents the power dissipated in all other circuit blocks of the transmitter and receiver to operate the communicating terminals, and $p_h$ denotes the power consumed by UAV in hovering, which takes into account the influence of air density, rotor disc area, blade angular velocity, rotor radius, rotor solidity, profile drag coefficient, incremental correction factor of induced power, and the weights of payloads. $p_{_{RIS}}$ denotes the RIS power consumption and can be given as follows: 

\begin{equation}
p_{_{RIS}}=N\left[P_r(b)+P_F\right],
\end{equation}
where $P_F$ and $P_r(b)$ are the power consumption of the diode in forward biased mode to operate in ON state\cite{RISONOFF} and phase resolution\cite{RISCONSUMPTION}, respectively. 
The parameter $b$ gives the phase resolution in number of bits, which depends on the RIS hardware complexity and power consumption. Since each element of RIS is genuinely assumed to be a phase shifter, the phase resolution is related to the phase shifting value of $2^b$ per RIS element. For example, the power consumption for 4-bit and 6-bit finite phase resolution is $P_n(4) = 45$mW and $P_n(6) = 78$mW, respectively, whereas the power consumption for infinite phase resolution is $P_n(\infty) = 32$W\cite{RISRESOLUTION}. 
Therefore, the hardware power consumption of RIS increases as the number of RIS elements and the phase resolution increase\cite{RISCONSUMPTION,RISRESOLUTION}. 

\section{Performance characterization and analysis}\label{sec:Performance}
In this section, we first characterize the statistical properties of the received SNR $\gamma$. Considering two cases with small and large values for $N$, we first obtain the exact distribution of $A$. Then, we derive the approximate expression of $A$ distribution based on the CLT approximation. 
Following that, we apply the statistical properties of $A$ presented above to provide an analysis of the proposed Fisher-Snedecor $ \mathcal{F}$ composite fading channel model, including the AC, EE, and OP. In particular, we derive theoretical bounds and compare them with the exact distribution of the average SNR, where it has been found that the lower bound achieves relatively high accuracy. Moreover, the upper bound of OP is proposed to evaluate the diversity order of channel statistics. 

\subsection{Distribution of characterization $A$}\label{omega}
Noting that ${h_{_{u,i}}}$ are i.i.d. Fisher-Snedecor $\mathcal{F}$ RVs.
The expected value of ${h_{_{u,i}}}$ can be derived \cite[Eq.~ (3.194.3)]{2007Table} as follows:
\begin{equation}\label{Ehi}
\mathbb{E}[{h_{_{u,i}}}]={\left(\frac{m_s\Omega_m}{m\Omega_s}\right)}^{\frac{1}{2}}\times\frac{B\left(m+\frac{1}{2},m_s-\frac{1}{2}\right)}{B(m,m_s)},
\end{equation}
while the mean power of ${h_{_{u,i}}}$ is expressed as follows: 
\begin{equation}\label{Ehi2}
\begin{aligned}
\mathbb{E}[{h^2_{_{u,i}}}]=\frac{m_s\Omega_m}{m\Omega_s}\times\frac{B({m+1},{m_s-1})}{B(m,m_s)}=\frac{m_s\Omega_m}{(m_s-1)\Omega_s}.
\end{aligned}
\end{equation}
Here, let $\Omega_s^2={m_s}/{(m_s-1)}$ and $\Omega_m^2={(m_s-1)}/{m_s}$, 
the mean power can be normallized as $\mathbb{E}[{h^2_{_{u,i}}}]=\Omega=1$.
In consequence, the variance of ${h_{_{u,i}}}$ can be expressed as $\mathrm{VAR}[{h_{_{u,i}}}]=\mathbb{E}[{h^2_{_{u,i}}}]-(\mathbb{E}[{h_{_{u,i}}}])^2$.
We then discuss the distribution of $A$ for the two cases of a small and large number of RIS elements $N$, respectively.

(1) \textit{A Small Number of $N$}:

Firstly, using \cite[3.389.2]{2007Table}, we can characterize the MGF of the Fisher-Snedecor $\mathcal{F}$ variable, denoted by $M_{h_{_{u,i}}}(t)$, as
follows:
\begin{equation}
\begin{aligned}
M_{h_{_{u,i}}}(t)&=
\int_{0}^{\infty}{\frac{2({\frac{m\Omega_s}{m_s\Omega_m}})^m{h_{_{u,i}}}^{2m-1}}{B(m,m_s)\left(1+\frac{m\Omega_s{h_{_{u,i}}}^2}{m_s\Omega_m}\right)^{m+m_s}}e^{-th_{_{u,i}}}}dh_{_{u,i}}\\
&=\frac{1}{\sqrt{\pi}\Gamma(m)\Gamma(m_s)}G^{3 1}_{1 3}\left(\frac{m_s\Omega_m}{4m\Omega_s}t^2 \middle|\, \begin{gathered}
1-m\\
m_s, 0, \frac12,
\end{gathered}\right)
\end{aligned}
\end{equation}
where $G^{m n}_{p q}\left(z \middle|\, \begin{gathered}
a_1,...,a_p\\
b_1,..., b_q
\end{gathered}\right)$ is the Meijer’s G-function \cite[Eq.~(9.301)]{2007Table}. 
Since $h_{_{u,i}}$ are i.i.d RVs, the MGF of $A$ can be obtained by the multiplication of the MGF of $h_{_{u,i}}$ as follows:
\begin{equation}\label{mgfA}
\begin{aligned}
M_{A}(t)&=
\left[\frac{1}{\sqrt{\pi}\Gamma(m)\Gamma(m_s)}G^{3 1}_{1 3}\left(\frac{m_s\Omega_m}{4m\Omega_s}t^2 \middle|\, \begin{gathered}
1-m\\
m_s, 0, \frac12.
\end{gathered}\right)\right]^N
\end{aligned}
\end{equation}
Then, let $t=j\omega$, the characteristic function of $A$, denoted by $\phi_A(\omega)$, can be obtained as follows:
\begin{equation}\label{characterA}
\begin{aligned}
\phi_A(\omega)=
\left[\frac{1}{\sqrt{\pi}\Gamma(m)\Gamma(m_s)}G^{3 1}_{1 3}\left(-\frac{m_s\Omega_m}{4m\Omega_s}\omega^2 \middle|\, \begin{gathered}
1-m\\
m_s, 0, \frac12
\end{gathered}\right)\right]^N.
\end{aligned}
\end{equation}
Based on the Gil-Pelaez inversion [16], 
the exact CDF of $A$ can be obtained as follows:

\begin{equation}\label{exactCDF}
\begin{aligned}
F_A(t)&=\frac12+\frac{1}{2\pi}\int_{0}^{\infty}{\left(\frac{e^{j\omega t}\phi_A(-\omega)-e^{-j\omega t}\phi_A(\omega)}{j\omega}\right)}d\omega\\
&=\frac{1}{2}+\frac1\pi\int_{0}^{\infty}{\Im\left(\frac{e^{-jwt}\phi_A(\omega)}{\omega}\right)}d\omega,
\end{aligned}
\end{equation}
where $\Im(w)$ is the imaginary part of $\omega \in \mathbb{C}$ and $j=\sqrt{-1}$.

(2) \textit{A Large Number of $N$}
     
When $N$ is particularly large, using Eq.~(\ref{exactCDF}) to describe the exact distribution of $A$ is challenging due to its high computational complexity. Based on the CLT approximation, we derive expressions to approximate the distribution of $A$. 
According to the CLT, for a sufficiently large number of RIS elements ($N\gg1$), $A$ becomes a RV converging to a Gaussian distribution with mean value $\mu_A$ and variance $\sigma_A^2$, where $\mu_A=N\mathbb{E}[{h_{_{u,i}}}]$, $\sigma_A^2=N\mathrm{VAR}[{h_{_{u,i}}}]$. 
Consequently, the PDF of $A$, denoted by $f_{A}^{CLT}(a)$, can be approximated as 
\begin{equation}\label{CLTPDF}
f_{A}^{CLT}(a)=\frac{1}{\sqrt{2\pi{\sigma_A^2}}} \text{exp}\left({-\frac{(a-\mu_A)^2}{2\sigma_A^2}}\right),
\end{equation}
and the CDF, denoted by $F_{A}^{CLT}(a)$, can be expressed as 
\begin{equation}\label{CLTCDF}
F_{A}^{CLT}(a)=1-Q\left({\frac{a-\mu_A}{\sigma_A}}\right),
\end{equation}
where $Q(\cdot)$ is the Gaussian $Q$-function \cite{wireless}. According to the above derivations and the SNR $\gamma$ in Eq.~(\ref{SNR}), the CDF of $\gamma$, denoted by $F_{\gamma}^{CLT}(t)$, can be expressed as follows:

\begin{equation}
\begin{aligned}
F_{\gamma}^{CLT}(t)&=F_{A}^{CLT}\left(\sqrt{\frac{PL_{\tiny{C}}PL_{\tiny{U}}}{\gamma_{_0}}t}\right)\\
&=1-Q\left(\frac{\sqrt{\frac{PL_{\tiny{C}}PL_{\tiny{U}}}{\gamma_{_0}}}t-\mu_A}{\sigma_A}\right)
\end{aligned}
\end{equation}
	\vspace{-10pt}
	
\begin{figure}[t] 
	\centering 
	\includegraphics[scale=0.52]{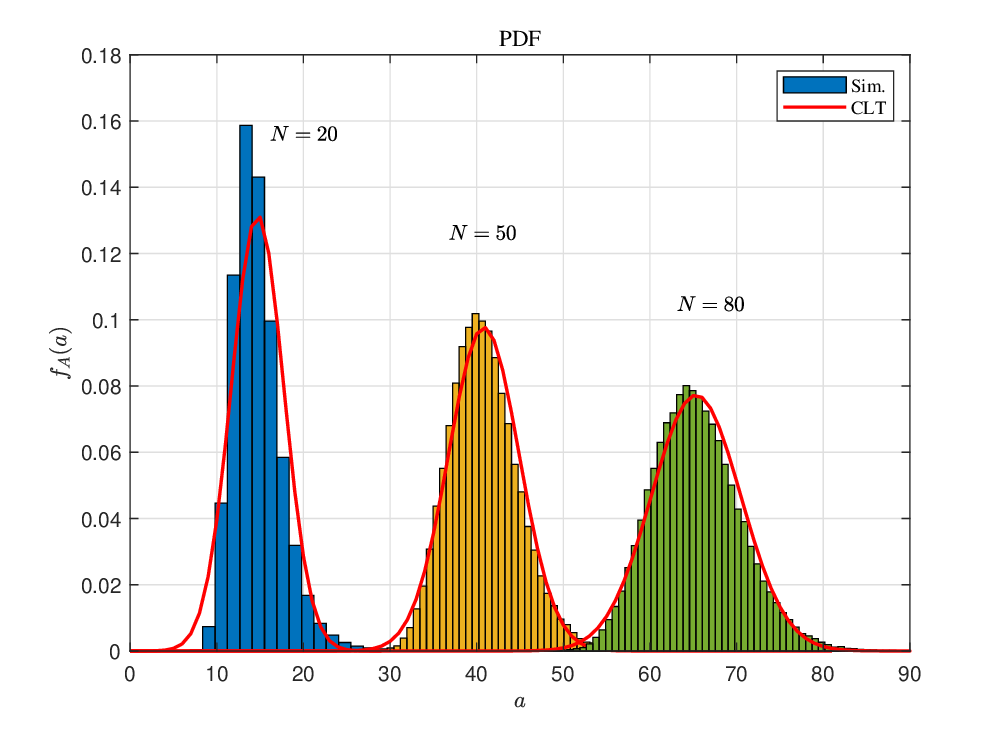}
	
	\vspace{-4pt}
	\caption{The comparison of the theoretical and simulated results of the PDF of $A$ under CLT approximation} 
	\vspace{-10pt}
	\label{Fig.PDF_CLT} 
\end{figure}
\begin{figure}[t] 
	\centering 
	\includegraphics[scale=0.52]{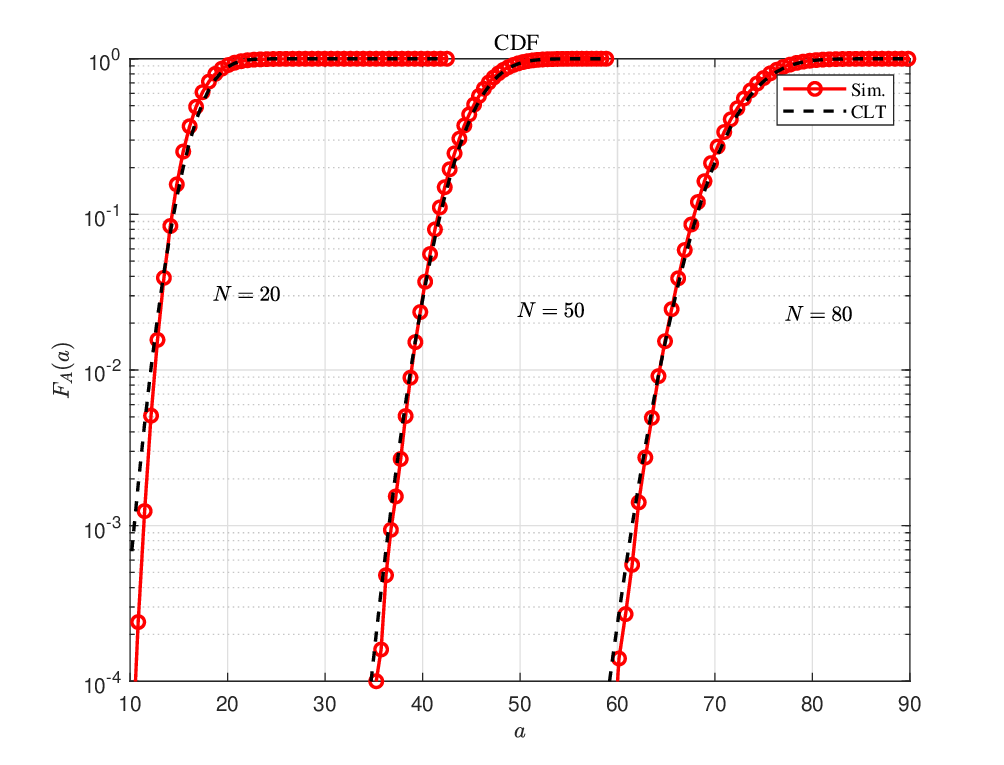}
	\vspace{-10pt}
	\caption{The comparison of the theoretical and simulated results of the CDF of $A$ under CLT approximation} 
	\vspace{-20pt}
	\label{Fig.CDF_CLT} 
\end{figure}

To verify the accuracy of the CLT approximation for $A$, we compare the theoretical curves with the exact distribution simulations in Figs.~\ref{Fig.PDF_CLT} and \ref{Fig.CDF_CLT}. 
Here, $N$ is set to $10$, $50$, and $100$, respectively, with $m=2$ and $m_s=2.5$. As illustrated in Figs.~\ref{Fig.PDF_CLT} and \ref{Fig.CDF_CLT}, the CLT approximation can better characterize the statistical performance of $A$ as $N$ increases. It is noteworthy that the CLT approximation is relatively accurate when $N$ exceeds 50 through simulation. 

\subsection{Average Capacity and Energy Efficiency}

The AC, denoted by $C$, 
can be evaluated as follows:
\begin{equation}
\begin{aligned}
C &= \mathbb{E}\left[ B\text{log}(1+\gamma)\right]\\
&=\int_{0}^{\infty}{ B\text{log}\left(1+\frac{\gamma_{_0}}{PL_{\tiny{C}}PL_{\tiny{U}}}A^2\right)f_A(a)}da.
\end{aligned}
\end{equation}
In general, it is hard to obtain the closed-form expression of $C$. 
The upper bound of the AC can be calculated using Jensen\mbox{'}s inequality as follows:
\begin{equation}
C \leq B\text{log}(1+\mathbb{E}[\gamma]).
\end{equation}
(1) \textit{A Small Number of $N$} 

The exact expectation of $A^2$ is equal to the second derivative of the MGF given in Eq.~(\ref{mgfA}), which can be expressed as follows:
\begin{equation}\label{exactE}
\mathbb{E}[A^2]=\left.\frac{d^2M_A(t)}{dt^2} \right|_{t=0}.
\end{equation}
The exact expectation of $A^2$ can be achieved by numerical calculations. Then the exact expectation of $\gamma$ can be obtained by applying the calculated numerical solution of $\mathbb{E}[A^2]$ into Eq.~(\ref{SNR}).
	
	\textit{Remark 1}: 
	This approach of numerical calculation will also significantly increase the amount of calculation when $N$ increases. 
	To simplify the calculation and gain more insights, we derive the relatively precise lower and upper bound expressions in the following. 
	
	\textit{Proposition 1}: The expectation of $\gamma$ can be lower-bounded and upper-bounded as follows: 
\begin{equation}\label{snrublb}
\begin{aligned}
{\frac{N^2\gamma_{_0}(m_s-1)}{{PL_{\tiny{C}}PL_{\tiny{U}}}m}}&\times\left[\frac{B\left(m+\frac{1}{2},m_s-\frac{1}{2}\right)}{B(m,m_s)}\right]^2 \\
&\le \mathbb{E}[\gamma]\le \frac{\gamma_{_0}N^2}{PL_{\tiny{C}}PL_{\tiny{U}}}.
\end{aligned}
\end{equation}

\textit{Proof}: 
First, we ultilize the complex Cauchy\mbox{-}Schwarz\mbox{-}Buniakowsky inequality\cite{2007Table}, and then $\mathbb{E}[A^2]$ can be upper-bounded as follows:
\begin{equation}\label{Eub}
\mathbb{E}\left[\sum_{i=1}^{N}{h_{_{u,i}}}\right]^2\le \mathbb{E}\left[N\sum_{i=1}^{N}{h^2_{_{u,i}}}|\right]= N\sum_{i=1}^{N}\mathbb{E}\left[{h_{_{u,i}}}^2\right],
\end{equation}
where $\mathbb{E}\left[{h_{_{u,i}}}^2\right]$ given in Eq.~(\ref{Ehi2}) is set to $1$ for channel normalization. 
Then, the lower bound of $A^2$ can be calculated according to Jensen\mbox{'}s inequality as
\begin{equation}\label{Elb}
\mathbb{E}\left[\sum_{i=1}^{N}{h_{_{u,i}}}\right]^2 \ge  \left(\sum_{i=1}^{N}\mathbb{E}\left[{h_{_{u,i}}}\right]\right)^2=\left(N\mathbb{E}\left[{h_{_{u,i}}}\right]\right)^2,
\end{equation}
where $\mathbb{E}[{h_{_{u,i}}}]$ is given in Eq.~(\ref{Ehi}). Substituting Eqs.~(\ref{Eub}) and (\ref{Elb}) into Eq.~(\ref{SNR}), respectively, the expectation of $\gamma$ can be lower-bounded and upper-bounded in Eq.~(\ref{snrublb}), which completes the proof. 


(2) \textit{A Large Number of $N$}

When the number of $N$ is large, by applying the binomial expansion theorem to Eq.~(\ref{SNR}), the expectation of $\gamma$ is given by 
\begin{equation}\label{SNRCLT}
\mathbb{E}[\gamma]=\frac{\gamma_{_0}\mathbb{E}[A^2]}{PL_{\tiny{C}}PL_{\tiny{U}}}= \frac{\gamma_{_0}({\mu_A^2}+{\sigma_A^2})}{PL_{\tiny{C}}PL_{\tiny{U}}}.
\end{equation}

Consequently, based on the definition of EE (which is defined as the ratio of the average capacity $C$ to the corresponding power consumption $P_{tot}$), we can derive the EE of the EWC system as follows:
\begin{equation}
EE=\frac{B\text{log}{(1+\frac{\gamma_{_0}\mathbb{E}[A^2]}{PL_{\tiny{C}}PL_{\tiny{U}}})}}{\mu Ps+NP_r(b)+p_c+p_h}.
\end{equation}
 \begin{figure}[t] 
	\centering 
	\includegraphics[scale=0.52]{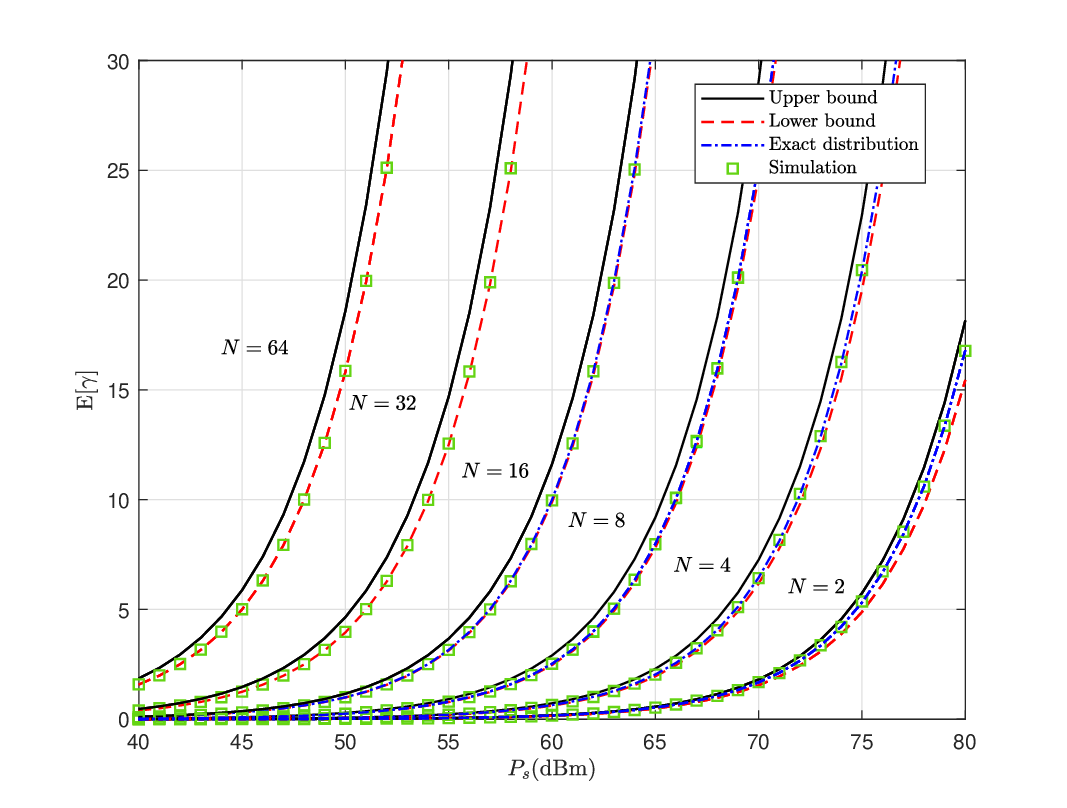}
	\vspace{-20pt}
	\caption{The comparison of average SNR between the exact distribution,  
		the upper bound, and the lower bound 
		with a small number of $N$.} 
	\vspace{-10pt}
	\label{Fig.SNR} 
\end{figure}
\begin{figure}[t] 
	\centering 
	\includegraphics[scale=0.52]{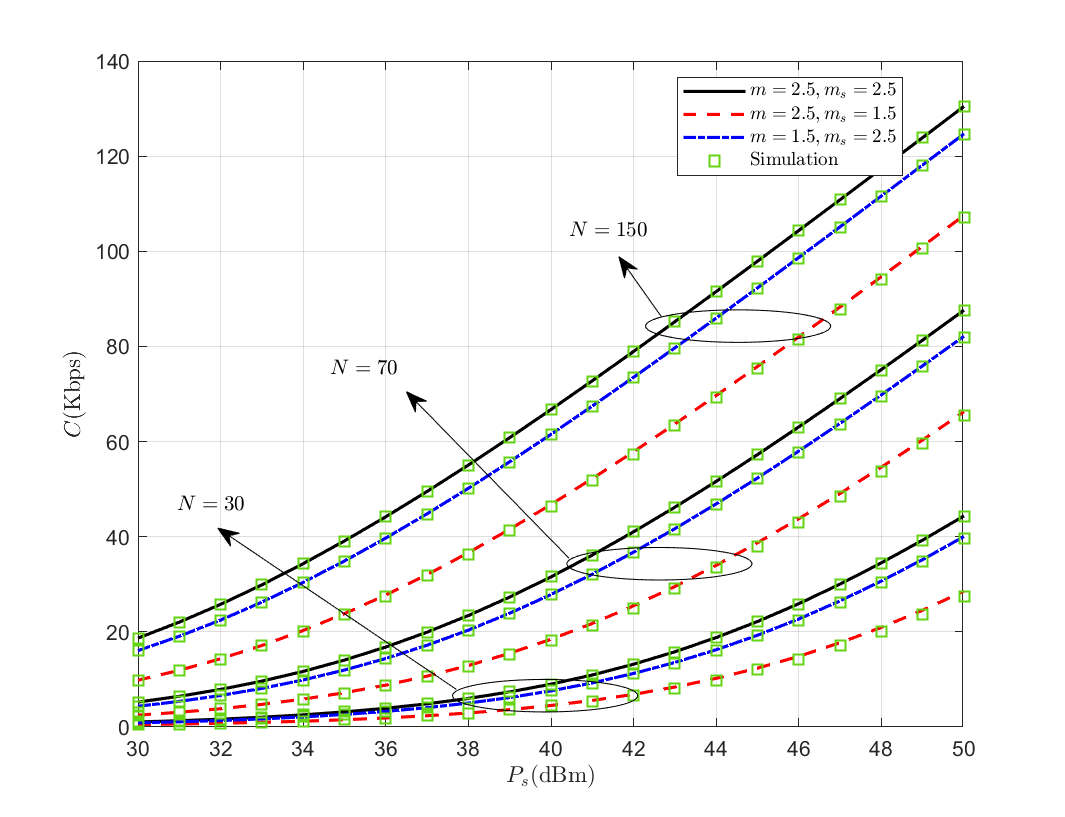}
	\vspace{-20pt}
	\caption{Average capacity corresponding to $N$, $m$, and $m_s$ under CLT approximation} 
	\vspace{-10pt}
	\label{Fig.AC_CLT} 
\end{figure}
	\vspace{-15pt}

Figure~\ref{Fig.SNR} shows our proposed bounds of $\mathbb{E}[\gamma]$ in Eq.~(\ref{snrublb}), the exact value of $\mathbb{E}[\gamma]$, and Monte-Carlo simulations. 
Compared with the upper bound, the lower bound of $\mathbb{E}[\gamma]$ is more accurate and basically tallies with the exact distribution as $N\ge8$. 
On the other hand, Fig.~\ref{Fig.AC_CLT} shows the AC corresponding to various $N$, $m$, and $m_s$ when the CLT approximation is taken into account for a large number of $N$. 
As illustrated in Fig.~\ref{Fig.AC_CLT}, the AC monotonically increases with $N$ for different $m$ and $m_s$, and then severely decreases as multipath fading and shadowing become more harsh. 

\textit{Remark 2}: It is obvious that emergency communication system designers can improve the AC by increasing the number of RIS elements $N$ in harsh environments, but this will also increase the power consumption of RIS, thus reducing the EE of the whole system. As a result, we will discuss how to achieve the balance between capacity and energy consumption when considering the optimization of EE.
\subsection{Outage Probability}
Based on the definition of OP and CLT approximation in Eq.~(\ref{CLTCDF}), the OP, denoted by $P_{out}^{CLT}$, can be re-expressed as follows:
\begin{equation}
\begin{aligned}
P_{out}^{CLT}&=Pr\left\{\gamma<\gamma_{th}\right\}=\int_{0}^{\gamma_{th}} f(\gamma) d\gamma\\
&=F_{\gamma}\left(\gamma_{th}\right)=F_{A}^{CLT}\left(\sqrt{\frac{PL_{\tiny{C}}PL_{\tiny{U}}}{\gamma_0}\gamma_{th}}\right).
\end{aligned}
\end{equation}
where $\gamma_{th}$ represents the desired SNR threshold. 

\textit{Remark 3}: The results of the CLT approximation in Eq.~(\ref{CLTPDF}) are not accurate for $a \rightarrow 0^{+}$. 
	As a result, the CLT-based channel statistics cannot be used to analyze the diversity order\cite{asy}. 
	
	To address the abovementioned issue, we further derive exact channel statistics for channel coeffience near 0 without using the CLT approximation, as shown in the following proposition.

\textit{Proposition 2}: When ${h_{u,i}} \rightarrow 0$, the upper bound of the OP, denoted by $P_{out}^{ub}({\gamma_0})$, is given as follows:
\begin{equation}
\begin{aligned}
P_{out}^{ub}({\gamma_0})
=\left[\frac{2\left({\frac{m\Omega_s}{m_s\Omega_m}}\right)^{m}\Gamma(2m)}{B(m,m_s)}\right]^N \frac{\left(\frac{\gamma_{th}}{{\gamma_0}}\right)^{Nm}}{2Nm\Gamma(2Nm)}.
\end{aligned}
\end{equation}

\textit{Proof}:
 See Appendix.

\begin{figure}[t] 
	\centering 
	\includegraphics[scale=0.53]{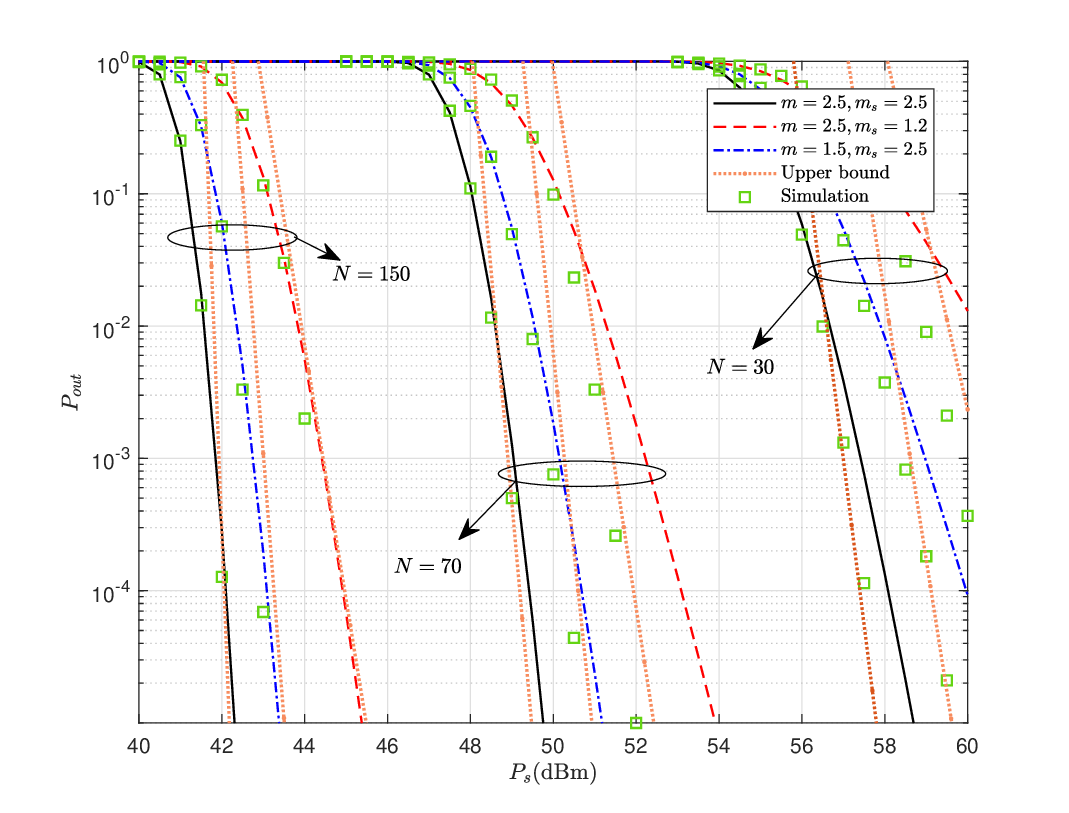}
	
	\vspace{-15pt}
	\caption{Outage probability corresponding to $N$, $m$, and $m_s$ under CLT approximation.}  
	\vspace{-10pt}
	\label{Fig.OP_CLT} 
\end{figure}

Figure~\ref{Fig.OP_CLT} depicts the OP versus $P_s$ with different numbers of RIS elements $N$ and fading parameters to compare the theoretical OP obtained via CLT approximation $P_{out}^{CLT}$, the upper bound $P_{out}^{ub}$ given by Proposition 2, and the numerical results. As shown in Fig.~\ref{Fig.OP_CLT}, although the CLT approximation results are relatively accurate for $N=70$ when experiencing moderate and harsh multipath fading, the results for both $N=30$ and $N=70$ are inaccurate under harsh shadowing. 
After that, the CLT approximation is relatively accurate for severe shadowing when $N=150$. It is indicated that a relatively larger number of $N$ is required to accurately fit the curves of OP when experiencing severe shadowing. Furthermore, all simulations gradually approach their respective asymptotic curves derived from Proposition 2, which validates our analyses. In general, emergency system designers can improve the OP by increasing the number of RIS elements, especially in harsh environments. 

\section{Optimization Problems}\label{sec:Optimization}
Due to scarce communication resources and tremendous rescue demand for the considered EWC system, new problems emerge in two aspects. 
Considering the rated load and prescriptive flight altitude of UAV, we first need to achieve a balance between the energy efficiency enhancements of the EWC system and the number of RIS reflecting elements equipped with UAV. Second, due to the emergency communication demands of multiple trapped users, we should maximize the coverage area while guaranteeing link reliability and the limited flight altitude of UAV. 
Thus, we formulate two optimization problems in this section as follows: one is to maximize system energy efficiency $EE$ subject to the constraint of RIS element number $N$, and the other is to maximize coverage area subject to the constraints of OP and flight altitude of the RIS-equipped UAV. 
\subsection{Energy Efficiency Maximization}
As the number of RIS elements is limited, we formulate the $EE$ maximization problem to achieve the balance between the rated load quality of UAV and the performance of RIS, which can be expressed as follows:
\begin{equation}
\begin{aligned}
\boldsymbol{P1}:&\max_{N}\quad \frac{B\text{log}{(1+\frac{P_s}{N_0}\frac{\mathbb{E}[A^2]}{PL_{\tiny{C}}PL_{\tiny{U}}})}}{\mu Ps+NP_r(b)+p_c+p_h}\\
&\begin{array}{r@{\quad}r@{}l@{\quad}l}
\rm{s. t.} &N_{min}\le N \le N_{max},
\end{array}
\end{aligned}
\end{equation}
\begin{figure*}[!b]
	\rule[-7pt]{18.07cm}{0.05em}  
	\begin{numcases}{\label{PP2}	{\boldsymbol{P2}}: \max_{N,r}~H(N)=}
	{B\text{log}{\left(1+\frac{\gamma_{_0}\left(N\mathbb{E}\left[{h_{_{u,i}}}\right]\right)^2}{PL_{\tiny{C}}PL_{\tiny{U}}}\right)}}-r\left(\mu Ps+NP_r(b)+p_c+p_h\right),  ~~~~~~~~~~~~~~\text{if}~ N\le N_{th};\tag{\ref{PP2}{a}}\vspace{1ex}\\
	{B\text{log}{\left(1+\frac{\gamma_{_0}(N^2\mathbb{E}\left[{h_{_{u,i}}}\right]^2+N\mathbb{D}[h_i])}{PL_{\tiny{C}}PL_{\tiny{U}}}\right)}}-r\left(\mu Ps+NP_r(b)+p_c+p_h\right), ~~\text{if}~ N> N_{th},\tag{\ref{PP2}{b}}\vspace{1ex}
	\end{numcases}
	\begin{equation}
	{\rm{s. t.}}~~~~~~~~~~~~~~N_{min}\le N\le N_{max}.\\	
	\tag*{(34c)}
	\setcounter{equation}{34}
	\end{equation}
\end{figure*}
where $N_{max}$ is the maximum number of RIS elements regarding the load capacity of the UAV and ${N_{min}}$ is the minimum number of RIS elements to accurately describe the proposed lower bound of $\mathbb{E}[\gamma]$ as shown in Fig.\ref{Fig.SNR}\footnote{ Generally, $N_{max}$ also depends on the boundary of the far field and the near field of the antenna array, which is defined as ${2D^2}/{\lambda}<d$ with $D$ denoting the largest dimension of the antenna. The size of one RIS element is typically in the range from ${\lambda}/{8} $ to $\lambda/4$\cite{RIS2DL}. As such, this limitation becomes more evident at low frequencies but is much smaller than the size of UAV or RIS at a frequency of 1.5 GHz under the EWC scenario. In the case of $N_{min}$, we only consider the more accurate lower bound for the following optimization problems for simplicity.}. 
To derive solutions in closed-forms, we set $N_{th}$ as the threshold between the small number case and the large number case. 
\footnote{ As shown in Fig.~6, the lower bound curves basically tallies with the exact distribution as $N\ge8$ for the case of a small number of $N$, so the threshold $N_{th}$ mainly depends on the case of a large number. As shown in Figs. 4 and 5, the PDF and CDF of $A$ under CLT approximation are relatively accurate when $N $ exceeds about 50 through simulation. To this end, the value of $N_{th}$ can be determined. }
For $N\le N_{th}$, we adopt the lower bound of $\mathbb{E}[A^2]$ in Eq.~(\ref{snrublb}) to calculate average capacity. Meanwhile, the CLT approximation in Eq.~(\ref{SNRCLT}) is used for $N> N_{th}$. 
  
In general, $\boldsymbol{P1}$ is a single-ratio fractional programming problem, which is not convex. Nevertheless, since $\frac{ \partial^2 C }{ \partial N^2 }< 0$ holds for the two cases, the average capacity is concave with respect to $N$. Accordingly, $\boldsymbol{P1}$ is in the form of a ratio of concave and convex functions with respect to $N$, which is known as the concave-convex fractional programming (FP) problem. Based on the Dinkelbach's Transform in \cite{DINKLE}, $\boldsymbol{P1}$ is reformulated as Eq.~(\ref{PP2}) with a new auxiliary variable $r$ as shown at the bottom of this page.
For a given $N$, the auxiliary variable $r$ in each iteration $t$, denoted by $r_t$, can be iteratively updated in closed-form as 
\begin{equation}\label{rt}
r_t=\frac{C(N_{t-1})}{P_{tot}(N_{t-1})}.
\end{equation}
Then, by solving the first derivatives of Eqs. ({\ref{PP2}{a}}) and ({\ref{PP2}{b}}) and making them equal to 0, the optimal value of $N_t$, denoted by $N_t^\ast$, can be obtained as follows:
	
	\begin{numcases}
	{N_t^\ast=\hspace{-1mm}}\label{deri1}\frac{\Theta}{R}+\sqrt{\frac{(\Lambda\Theta^2-R^2)}{\Lambda R^2}},~~~~~~~~~~~~~
	\text{if}~N<N_{th};\tag{\ref{deri1}{a}}\vspace{1ex}\\%
	\frac{\left(\Lambda \Theta- \Xi R+\hspace{-1mm}\sqrt{(\Lambda \Theta \hspace{-1mm}-\Xi R)^2\hspace{-1mm}-4\Lambda R(R-\Theta\Xi)}\right)}{2\Lambda R}, \nonumber \\
	~~~~~~~~~~~~~~~~~~~~~~~~~~~~~~~~~~~~~~~ \text{if}~N\ge N_{th}.\tag{\ref{deri1}{b}}\vspace{1ex}
	\setcounter{equation}{36}
	\end{numcases}
	where $\Lambda=\frac{\gamma_0\mathbb{E}[{h_{_{u,i}}}]^2}{PL_{\tiny{C}}PL_{\tiny{U}}}$, $\Xi=\frac{\gamma_0\mathbb{D}[{h_{_{u,i}}}]}{PL_{\tiny{C}}PL_{\tiny{U}}}$, $\Theta=\frac{2B}{\text{ln}2}$, \vspace{1ex} and $R=r_{t-1}P_r(b)$.
Since the second derivatives of Eqs.~(34a) and (34b) can be calculated as follows:

\begin{subnumcases}
{\frac{ \partial ^2 H(N) }{ \partial N^2}=}\frac{\Theta\Lambda\left(1-\Lambda N^2\right)}{\left(1+\Lambda N^2\right)^2},  ~~~~~~~~~~~~~\text{if}~N<N_{th}; \tag*{(37a)}\\
-\frac{\Theta\left[\Lambda \left(\Lambda N^2-1\right)+\Xi ^2(2\Lambda N/\Xi +1)\right]}{\left(1+\Lambda N^2+\Xi N\right)^2}, \nonumber\\
	~~~~~~~~~~~~~~~~~~~~~~~~~~~~~~~~~\text{if}~N\ge N_{th},\tag*{(37b)}
\end{subnumcases}	\setcounter{equation}{37} 
and for both $N<N_{th}$ and $N\ge N_{th}$, $\frac{ \partial ^2 H(N^\ast) }{ \partial N^2}< 0$ can be obtained by substituting Eqs.~(36a) and (36b) into Eqs.~(37a) and (37b), respectively, so $N^\ast$ in Eq.~(36) is the maximum point.
In the following, we provide the details of the proposed algorithm in \textbf{Algorithm 1}. 
The overall iterative algorithm is proved to converge to the global optimal solution in \cite{DINKLE}. 

\begin{algorithm}[htb] 
	\caption{ Dinkelbach's Transform for Optimization of $N$  } 
	\label{Dinkelbach} 
	\begin{algorithmic}[1] 
		\STATE Initialize $t=1$, $r_1=0$, $N_1^\ast\in [N_{th}, N_{max}]$, set $N_{th}=50$, maximum iterations $t_{max}=5000$, error tolerance $\epsilon=10^{-6}$.\
			
		\STATE 	Set $t=2$, $r_{2}^\ast=\frac{C(N_1^\ast)}{P_{total}(N_1^\ast)}$.\

		\WHILE {$t<t_{max}$ \textbf{and} $|r_t^\ast-r_{t-1}^\ast|\ge \epsilon$}

		\STATE  Update $N_t^\ast$ for fixed $r_t$ according to Eq.~(\ref{deri1}{b}). \
		\STATE Update $r_t^\ast$ using Eq.~(\ref{rt}).\
		\STATE Set $t=t+1$.\
		\ENDWHILE
		\IF {$N_t^\ast\in [N_{min}, N_{th}]$}
		\STATE 	Set $r_2^\ast=r_t^\ast$, $t=2$.\
		\WHILE {$t<t_{max}$ \textbf{and} $|r_t-r_{t-1}|\ge \epsilon$}
		\STATE Update $N_t^\ast$ for fixed $r_t$ according to Eq.~(\ref{deri1}{a}). \
		\STATE Update $r_t^\ast$ using Eq.~(\ref{rt}).\
		\STATE Set $t=t+1$.\
		\ENDWHILE
		\IF {$N_t^\ast > N_{th}$}
		\STATE $N_t^\ast=N_{th}$.
		\ENDIF
			\ENDIF
		\STATE  By substituting $\lceil N_t^\ast\rceil$ and $\lfloor N_t^\ast \rfloor$ in to $\textbf{P2}$, $N_t^\ast$ is selected as the one that makes $H(N)$ larger. 
	\end{algorithmic}
\end{algorithm}

\subsection{Coverage Area Maximization}
In this subsection, we aim to deploy the RIS-equipped UAV at an optimal flight altitude to serve as many UEs as possible while ensuring link reliability. In particular, link reliability is generally evaluated using the OP. 
The UEs can only successfully communicate with the command center as $P_{out}<P^{th}_{out}$, where $P^{th}_{out}$ denotes the threshold outage probability.
Therefore, we formulate $\boldsymbol{P3}$ to derive the maximum value of $r_u$ subject to the constraints of the OP and flight altitude, which can be expressed as follows:

\begin{subequations}\label{P3}
\begin{align}
{\boldsymbol{P3}}:\max_{h, r}\quad &r_u \\
\rm{s. t.}\quad
&1-Q\left({\frac{\sqrt{\frac{PL_{\tiny{C}}PL_{\tiny{U}}}{\gamma_{_0}}\gamma_{th}}-\mu_A}{\sigma_A}}\right)\le P^{th}_{out};\\
&h_{min}\le h\le h_{max}.
\end{align}
\end{subequations}
where $h_{min}$ and $h_{max}$ are the minimum and maximum values of the flight altitude defined by aviation authorities, respectively. 
The Q-function in Eq.~(\ref{P3}{b})
can be converted into 
$\left(\sqrt{\frac{PL_{\tiny{C}}PL_{\tiny{U}}\gamma_{th}}{\gamma_{_0}}}-\right.\\ \left.\mu_A\right)/{\sigma_A}\le \xi$, where $\xi$ is the parameter according to the error function table\cite{Specialerf}. Based on the trigonometric relation $h = r_u \text{tan}\theta$ and substituting Eqs.~(\ref{plm1}) and (\ref{plmodel}) into the logarithm of Eq.~(\ref{P3}{b}), Eqs.~(\ref{P3}{b}) and (\ref{P3}{c}) can be reformulated as

\begin{numcases}{\label{gfunc}}
g_1(r_u,\theta)=5{\alpha}\text{lg}(z_c^2+r_u^2\text{tan}^2\theta)+20\text{lg}(r_u\text{sec}\theta\nonumber\\
~~~~~~~~~~~+\frac{(\eta_{_{LoS}}-\eta_{_{NLoS}})}{1+ae^{-b(\theta-a)}}+M\le 0; \tag{\ref{gfunc}{a}}\vspace{1ex}\\
g_2(r_u, \theta)=-r_u\text{tan}\theta+h_{min}\le 0;\tag{\ref{gfunc}{b}}\vspace{1ex}\\
g_3(r_u, \theta)= r_u\text{tan}\theta-h_{max}\le 0, \tag{\ref{gfunc}{c}}
\setcounter{equation}{39}
\end{numcases}
where $M=20\text{lg}{\frac{4{\pi}}{\lambda}}+\eta_{NLoS}-10\text{lg}\left(\frac{\xi \sigma_A+ \mu_A}{k}\right)^2$ and  $k={\gamma_{th}}/{\gamma_{_0}}$.
\setcounter{equation}{37}

 The Hessian Matrix of $g_1(r_u, \theta)$, denoted by $H(g_1)$, satisfies $|H(g_1)|> 0$ by numerical calculation. However, $|H_{g_2}(r_u, \theta)|$ and $|H_{g_3}(r_u, \theta)|$ are both less than 0. 
In light of this, ${\boldsymbol{P3}}$ is a non-convex optimization problem, and the KKT condition can lead to its suboptimal solution. Consequently, ${\boldsymbol{P3}}$ can be reformulated as follows:
\begin{align}\label{P4}
{\boldsymbol{P4}}:\min_{h, r}\quad &-r_u \tag*{(40)}\\
\rm{s. t.}\quad
&g_i(r_u, \theta)\le 0, i=1,2,3.\nonumber
\setcounter{equation}{40}
\end{align}
Thus, the Lagrangian function, denoted by $L$, is given by
\begin{equation}
\begin{aligned}
L&= -r_u+\omega_1g_1(r_u,\theta)+\omega_2g_2(r_u, \theta)+\omega_3g_3(r_u, \theta),
\end{aligned}
\end{equation}
where $\omega_1$, $\omega_2$, and $\omega_3$ denote the Lagrange multipliers.  After that, ${\boldsymbol{P4}}$ can be solved by KKT condition to derive the optimal solution, 
which can be described as follows:
\begin{numcases}{\label{PP3}}
\frac{ \partial L}{ \partial r_u}= -1+\omega_1\frac{ \partial g_1}{ \partial r_u}-\omega_2\text{tan}\theta+\omega_3\text{tan}\theta=0; \tag{\ref{PP3}{a}}\vspace{1ex}\\
\frac{ \partial L}{ \partial \theta}= \omega_1\frac{ \partial g_1}{ \partial \theta}-\omega_2r_u\text{sec}^2\theta^\ast+\omega_3r_u\text{sec}^2\theta=0; \tag{\ref{PP3}{b}}\vspace{1ex} \\
\omega_ig_i(r_u, \theta)=0,~i=1,2,3;  \tag{\ref{PP3}{c}}\vspace{1ex} \\
\omega_1\ge 0,\omega_2\ge 0,\omega_3\ge 0;  \vspace{1ex}\tag{\ref{PP3}{d}} \\
g_i(r_u,\theta)\le 0,~i=1,2,3, \tag{\ref{PP3}{e}}\vspace{1ex}
\end{numcases}
\setcounter{equation}{42}
where ${ \partial g_1 }/{ \partial r_u}$ and ${ \partial g_1}/{ \partial \theta}$ can be calculated as follows:
\begin{numcases}{\label{g1deri}}
\frac{ \partial g_1 }{ \partial r_u}=\frac{10\alpha}{\text{ln}10}\frac{r_u\text{tan}^2\theta }{(z_c^2+r_u^2\text{tan}^2\theta)}+\frac{20}{\text{ln}10r_u};\tag{\ref{g1deri}{a}}\vspace{1ex}\\ 
\frac{ \partial g_1}{ \partial \theta}= \frac{5\alpha}{\text{ln}10}\frac{2r_u^2\text{tan}\theta\text{sec}^2\theta}{z_c^2+r_u^2\text{tan}^2\theta}\nonumber\\
~~~~~+\frac{(\eta_{_{LoS}}-\eta_{_{NLoS}})abe^{-b\theta+ab}}{(1+ae^{-b\theta+ab})^2}+\frac{20\text{tan}\theta}{\text{ln}10}.\tag{\ref{g1deri}{b}}\vspace{1ex}
\end{numcases}
\setcounter{equation}{43}
Based on the above analyses, a general form of the optimal $\theta^\ast$ and $r_u^\ast$ can be derived according to the following three cases.
 

\underline{\textbf{Case 1}}: If $h_{min} < r_u^\ast \text{tan}\theta^\ast < h_{max}$, then $\omega_2=\omega_3=0$ according to Eqs.~(\ref{PP3}{c}) and (\ref{PP3}{d}). Hence, Eq.~(\ref{PP3}) can be rewritten as follows:
\begin{numcases}{\label{case1}}
-1+\omega_1\frac{\partial g_1}{\partial r_u^\ast}=0; \tag{\ref{case1}{a}}\vspace{1ex}\\
\omega_1\frac{ \partial g_1}{ \partial \theta^\ast}=0; \tag{\ref{case1}{b}}\vspace{1ex}\\
\omega_1g_1(r_u^\ast,\theta^\ast)= 0. \tag{\ref{case1}{c}}\vspace{1ex}
\end{numcases}\setcounter{equation}{44}
The value in Eq.~(\ref{g1deri}{a}) is positive obviously. Therefore, Eq.~(\ref{case1}{a}) states that $\omega_1=(\frac{\partial g_1}{\partial r_u})^{_{-1}}>0$\vspace{1ex}, which is consistent with Eq.~(\ref{PP3}{d}). 
  Since Eqs.~(\ref{g1deri}{a}) and (\ref{g1deri}{b}) contain transcendental functions such as the exponential and trigonometric functions, it is quite challenging to find analytical solutions to these complex transcendental equations. 
Therefore, we adopt the ``fsolve" function in MATLAB, which provide a Newton-like algorithm called 'trust-region-dogleg' to solve the transcendental equations. 

\underline{\textbf{Case 2}}: If $h_{min}=r_u^\ast \text{tan}\theta^\ast$, then $\omega_2>0, \omega_3=0$ is reached according to Eqs.~(\ref{PP3}{c}) and (\ref{PP3}{d}). Hence, Eq.~(\ref{PP3}) can be rewritten as follows:
\begin{numcases}{\label{case2}}
-1+\omega_1\frac{\partial g_1}{\partial r_u^\ast}-\omega_2\text{tan}\theta^\ast=0; \tag{\ref{case2}{a}}\vspace{1ex}\\
\omega_1\frac{ \partial g_1}{ \partial \theta^\ast}-\omega_2r_u^\ast\text{sec}^2\theta^\ast=0;\tag{\ref{case2}{b}}\vspace{1ex}\\
\omega_1g_1(r_u^\ast,\theta^\ast)= 0; \tag{\ref{case2}{c}}\vspace{1ex}\\
\omega_2(-r_u^\ast\text{tan}\theta^\ast+h_{min})=0.\tag{\ref{case2}{d}}\vspace{1ex}
\end{numcases}\setcounter{equation}{45}
In specific, based on Eqs.~(\ref{case2}{a}) and (\ref{case2}{b}), \vspace{1ex}$\omega_1$ can be obtained as $\omega_1=(\omega_2\text{tan}\theta^\ast+1)(\frac{ \partial g_1}{ \partial r_u^\ast})^{-1}=\omega_2r_u\text{sec}^2\theta^\ast(\frac{ \partial g_1}{ \partial \theta^\ast})^{-1}$,\vspace{1ex} which reveals that $\omega_1$ and $\omega_2$ are both positive/negative.  Then, $\omega_1, \omega_2, r_u^\ast$, and $\theta^\ast$ can be calculated by solving the aforementioned simultaneous equations. Therefore, we can obtain the optimal  $\theta^\ast$ and $r_u^\ast$ by selecting solutions that satisfy both $\omega_ 1>0$ and $\omega_2>0$. 

\underline{\textbf{Case 3}}: If $h_{max}=r_u^\ast \text{tan}\theta^\ast$, then $\omega_2=0, \omega_3>0$ is reached according to Eqs.~(\ref{PP3}{c}) and (\ref{PP3}{d}). Hence, Eq.~(\ref{PP3}) can be rewritten as follows:
\begin{numcases}{\label{case3}}
-1+\omega_1\frac{\partial g_1}{\partial r_u^\ast}+\omega_3\text{tan}\theta^\ast=0; \tag{\ref{case3}{a}}\vspace{1ex}\\
\omega_1\frac{ \partial g_1}{ \partial \theta^\ast}+\omega_3r_u^\ast\text{sec}^2\theta^\ast=0;\tag{\ref{case3}{b}}\vspace{1ex}\\
\omega_1g_1(r_u^\ast,\theta^\ast)= 0; \tag{\ref{case3}{c}}\vspace{1ex}\\
\omega_3(-r_u^\ast\text{tan}\theta^\ast+h_{min})=0.\tag{\ref{case3}{d}}\vspace{1ex}
\end{numcases}\setcounter{equation}{46}
According to Eq.~(\ref{case3}{a}), \vspace{1ex}$\omega_1$ can be expressed as  $\omega_1=(1-\omega_3\text{tan}\theta^\ast)(\frac{ \partial g_1 }{ \partial r_u^\ast})^{-1}$, \vspace{1ex}meaning  that $\omega_1, \omega_3$ are simultaneously positive only for $\omega_3\text{tan}\theta^\ast<1$. Similar to the above analyses, the optimal $\theta^\ast$ and $r_u^\ast$ can be achieved by solving the aforementioned simultaneous equations after choosing the positive $\omega_1$ and $\omega_3$.

\begin{table*}[t]
	\caption{{Simulation parameters}\label{table1}}  
	\begin{center}
		\begin{tabular}{|c|c|c|}  
			\hline  
			& & \\[-6pt]  
			Parameter & Value & Definition\\ 
			\hline
			& & \\[-6pt]  
			$B$&20MHz&The available bandwidth\\
			\hline
			& & \\[-6pt]  
			$f$&1.5GHz&The frequency of communication\\
			\hline
			& & \\[-6pt]  
			$\eta_{_{LoS}}$&0.1dB&The excessive path loss of LoS propagation\\ 
			\hline
			& & \\[-6pt]  
			$\eta_{_{NLoS}}$&20dB&The excessive path loss of NLoS propagation\\ 
			\hline
			& & \\[-6pt]  
			$a,b$&4.88, 0.4472&The S-curve parameter specific to the environment \\ 
			\hline
			& & \\[-6pt]  
			$\alpha$&2&The path loss exponent\\ 
			\hline
			& & \\[-6pt]  
			$v$&1.1&The efficiency of the transmit
			power amplifier\\ 
			\hline
			& & \\[-6pt]  
			$p_h$&1.5kW&The power consumed by UAV in
			hovering\\
			\hline
			& & \\[-6pt]  
			$p_c$&50W&The power dissipated in all other circuit blocks of the transmitter and receiver\\
			\hline
			& & \\[-6pt]  
			$P_F$&1mW&The power consumption of diode in forward biased mode to operate in ON state\\
			\hline
			& & \\[-6pt]  
			$P_r(b)$&{ 78mW$\sim$0.15W}&The power consumption of phase resolution \\	
			\hline
			& & \\[-6pt]  
			$N_{min}, N_{max}$&8, 1000& The minimum and maximum number of RIS elements equipped by UAV\\	
			\hline
			& & \\[-6pt]  
			$\gamma_{th}$&10dB&The desired SNR threshold associated with $P_{out}$\\
			\hline
			& & \\[-6pt]  
			$P^{th}_{out}$&1e-4&The desired outage probability threshold\\		
			\hline
			& & \\[-6pt]  
			$h_{min}, h_{max}$&100m, 2km&The restrictions on the minimum and maximum flight altitudes. \\	
			\hline
		\end{tabular}
	\end{center}
\end{table*}
	\vspace{-10pt}

\section{Numerical results and discussions}\label{sec:Numerical}

In this section, we evaluate energy efficiency and coverage area under the EWC system with the modified Fisher-Snedecor $\mathcal{F}$ fading channel, respectively. 
Other parameter settings in the simulations are shown in TABLE \ref{table1} unless specified otherwise.

\begin{figure}[htbp] 
	\vspace{-10pt}
	\centering 
	\includegraphics[scale=0.6]{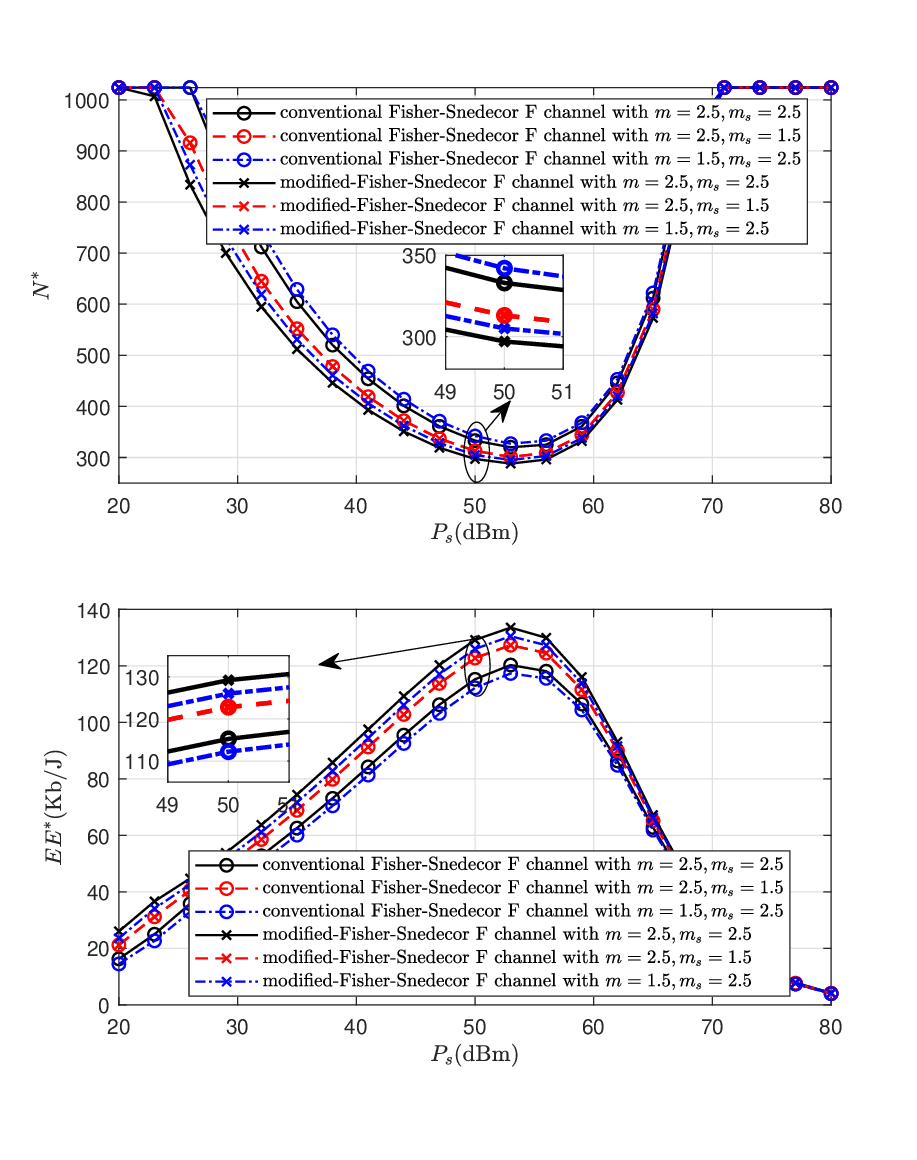}
	\vspace{-45pt}
	\caption{ The comparison of optimal $N^\ast$ and  $EE^\ast$ between the conventional $\mathcal{F}$ fading channel and the modified-$\mathcal{F}$ fading channel model with  different fading parameters $m$ and $m_s$.} 
	\vspace{-10pt}
	\label{Fig.Nopt_comparison} 
\end{figure} 

Figure~\ref{Fig.Nopt_comparison} shows the comparison of the optimal number of RIS elements $N^\ast$ and energy efficiency $EE^\ast$ between the modified-Fisher-Snedecor $\mathcal{F}$ fading channel given by \textbf{Algorithm \ref{Dinkelbach}} and the conventional Fisher-Snedecor $\mathcal{F}$ fading channel with different values of fading parameters $m$ and $m_s$. Here, we normalize the mean power as $\Omega=1$.{\footnote{
For the conventional $\mathcal{F}$ model, the mean power with $m_s=1.5$ and $m=2.5$ is taken as the normalization factor under all other parameter settings. Under this circumstance, the mean power of the conventional $\mathcal{F}$ model will become 1, which is consistent with modified-$\mathcal{F}$ fading channel. Then the mistakes of the conventional $\mathcal{F}$ channel can be clearly observed under different shadowing parameters $m_s$. 
	}}
	 First, we consider the impacts of increasing transmit power $P_s$ with $m = 2.5$ and $m_s = 2.5$, meaning that the channel experiences moderate fading and shadowing. It is clear that the modified-Fisher-Snedecor $\mathcal{F}$ fading channel outperforms the conventional Fisher-Snedecor $\mathcal{F}$ fading channel with normalized mean power. 
	 When $P_s$ reaches its lowest point at around $P_s=$ 53dBm, the optimal $N^\ast$ of the modified-$\mathcal{F}$ channel stops decreasing and then starts to rise as $P_s$ keeps increasing. Meanwhile, with the increase of $P_s$, the optimal $EE^\ast$ initially increases and reaches its highest point at the same value of $P_s=$ 53dBm, then drops. This is because the average capacity is more dominant than power consumption before reaching the extreme point. However, this impact will reverse as $P_s$ increases after reaching the extreme point. 
	 Then, we set $m_s=1.5$ and $m=1.5$. It is evident that as fading and shadowing become more harsh, the modified-$\mathcal{F}$ fading channel with \textbf{Algorithm \ref{Dinkelbach}} can achieve a lower optimal $EE^\ast$ for a larger $N^\ast$.  
	 This is because the deterioration of channel condition necessitates a greater number of RIS elements $N^\ast$, which also reduces $EE^\ast$. 
	 However, although the $EE^\ast$ of the conventional $\mathcal{F}$ fading channel decreases as $m$ decreases, it increases as $m_s$ decreases, which is unreasonable for channel conditions with harsh shadowing. 
	 	 
	 \begin{figure}[htbp] 
	 	\vspace{-10pt}
	 	\centering 
	 	\includegraphics[scale=0.62]{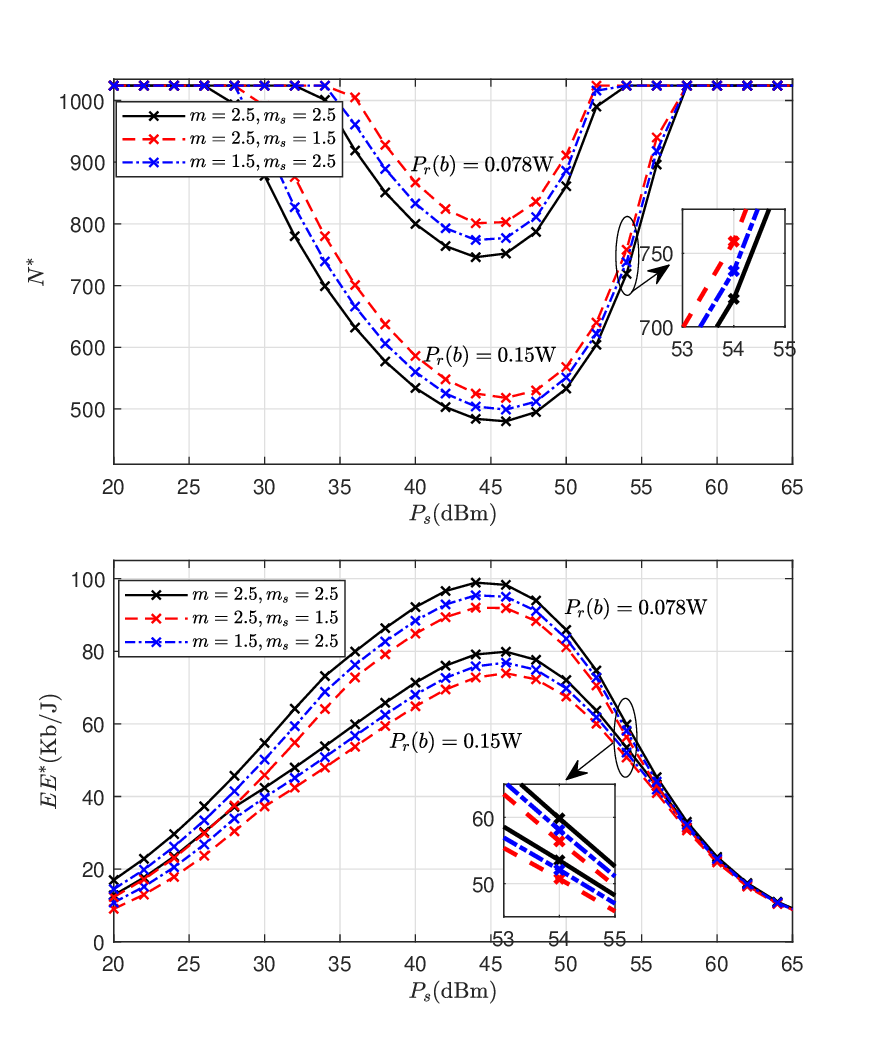}
	 	\vspace{-15pt}
	 	\caption{The optimal $N^\ast$ and  $EE^\ast$ versus $P_s$ corresponding to different phase resolutions $P_r(b)$ and fading parameters $m$, $m_s$.} 
	 	\vspace{-10pt}
	 	\label{Fig.Nopt_pu} 
	 \end{figure} 
	 \begin{figure}[t] 
	 	\vspace{-10pt}
	 	\centering 
	 	\includegraphics[scale=0.63]{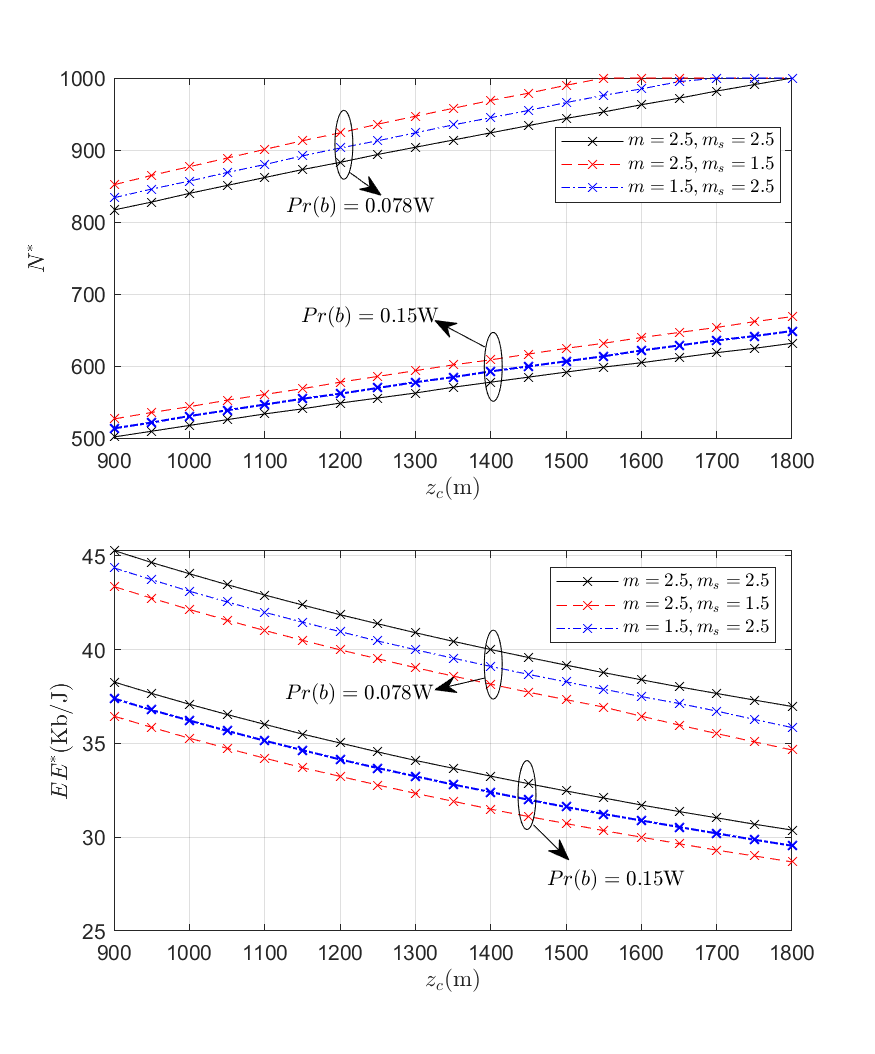}
	 	
	 	\vspace{-20pt}
	 	\caption{The optimal $N^\ast$ and $EE^\ast$ versus $z_c$ corresponding to different phase resolutions $P_r(b)$ and fading parameters $m$, $m_s$.} 
	 	\vspace{-10pt}
	 	\label{Fig.Nopt_zu} 
	 \end{figure} 
Then, Fig.~\ref{Fig.Nopt_pu} depicts the performances of optimal $N^\ast$ and $EE^\ast$ corresponding to different power consumption of phase resolution $P_r(b)$ versus $P_s$. As shown in Fig.~\ref{Fig.Nopt_pu}, when  $P_r(b)$ changes from 0.078W to 0.15W, the corresponding $EE^\ast$ decreases as the demanded optimal $N^\ast$ decreases due to the higher power consumption of each RIS element.  

Figure~\ref{Fig.Nopt_zu} shows the impacts of both the power consumption of phase resolution $P_r(b)$ and fading parameters on the optimal $N^\ast$ and $EE^\ast$ versus $z_c$. 
Likewise, $P_r(b)$ is set to $0.078$W and $0.15$W, respectively. It can be seen from Fig.~\ref{Fig.Nopt_zu} that $N^\ast$ monotonically increases as the distance between the command center and RIS $z_c$ increases for both $P_r(b)$=0.078W and 0.15W, while $EE^\ast$ keeps decreasing. That is, a greater optimal $EE^\ast$ can be attained for a smaller $N^\ast$ when $z_c$ is small. 
Similarly, $N^\ast$ becomes larger and $EE^\ast$ becomes smaller when the channel experiences more serious fading and shadowing, which is consistent with our previous derivation results. 
Moreover, both the optimal $N^\ast$ and $EE^\ast$ decrease as $P_r(b)$ changes from 0.078W to 0.15W, revealing that $P_r(b)$ dominates the optimization of $EE^\ast$. 

\begin{figure}[t] 
		\vspace{-10pt}
	\centering 
	\includegraphics[scale=0.67]{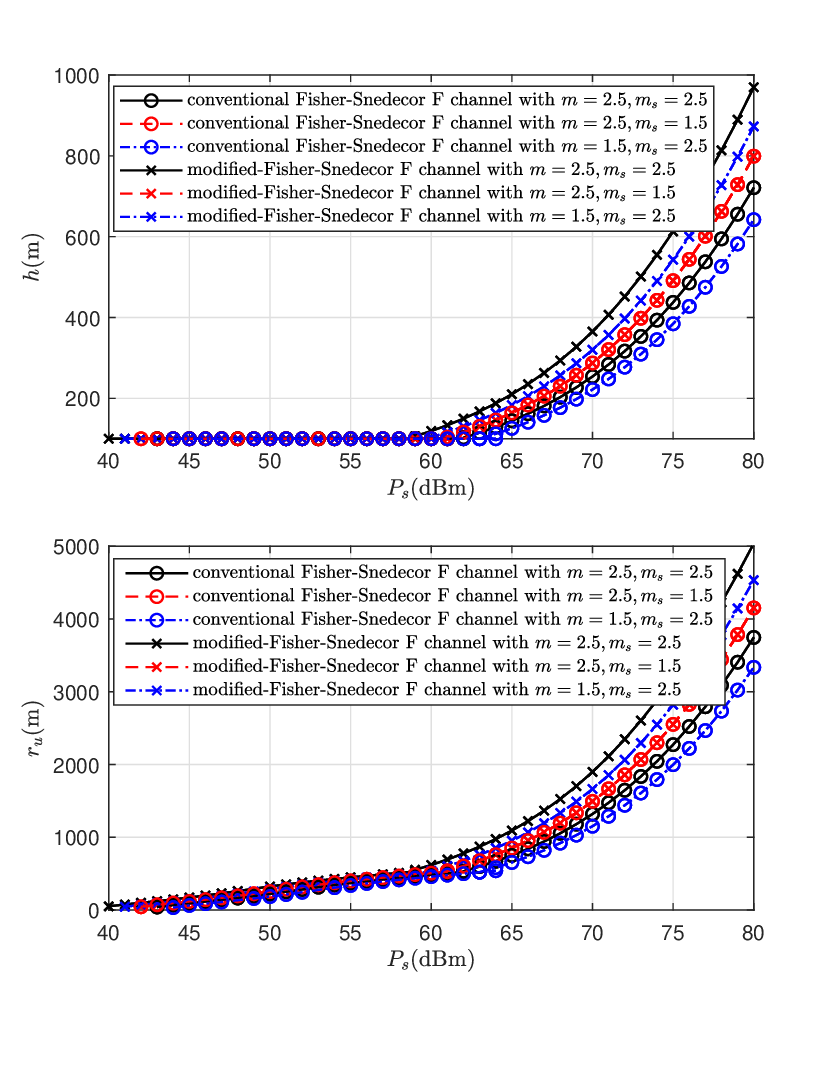}
	
	\vspace{-35pt}
	\caption{The comparison of optimal $h$ and $r_u$ between the conventional $\mathcal{F}$ fading channel and the modified-$\mathcal{F}$ fading channel model versus $P_s$ under different fading parameters $m$ and $m_s$.}
	\vspace{-20pt}
	\label{Fig.hopt_ps} 
\end{figure}
\begin{figure}[t] 
     \vspace{-10pt}
	\centering 
	\includegraphics[scale=0.68]{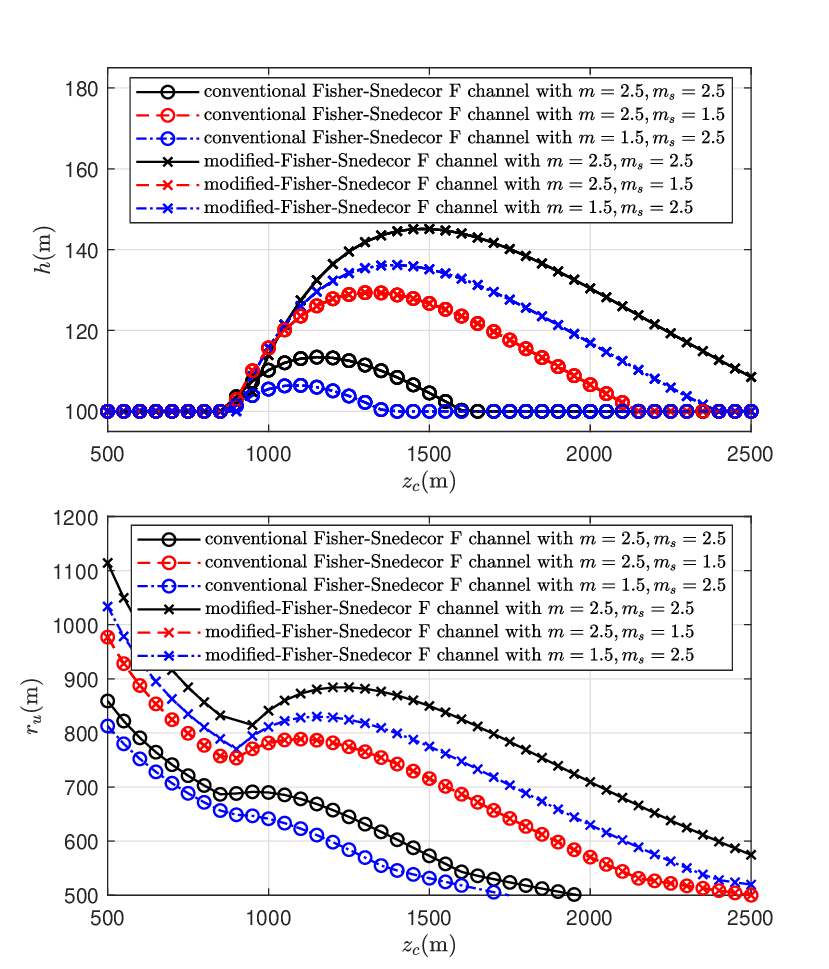}
	
     \vspace{-10pt}
	\caption{The comparison of optimal $h$ and $r_u$ between the conventional $\mathcal{F}$ fading channel and the modified-$\mathcal{F}$ fading channel model versus $z_u$ under different fading parameters $m$ and $m_s$.}
	\vspace{-10pt}
	\label{Fig.hopt_zu} 
\end{figure}
Figure~\ref{Fig.hopt_ps} compares the optimal flight altitude $h^\ast$ and service radius $r_u^\ast$ of the modified-Fisher-Snedecor $\mathcal{F}$ fading channel with those of the conventional Fisher-Snedecor $\mathcal{F}$ fading channel under different fading parameters. First, we focus on the impacts of increasing $P_s$ with both $m$ and $m_s$ being set as $2.5$. 
As shown in Fig.~\ref{Fig.hopt_ps}, the optimal value of $h^\ast$ corresponding to the modified-$\mathcal{F}$ fading channel remains constant at the minimum flight altitude 100m when $P_s\le$ 58dB, indicating that the solutions are derived according to Case 2. Correspondingly, the optimal value of $r_u^\ast$ slowly increases as $P_s$ increases. Then, the solutions are given by Case 1, in which both the optimal values of $h^\ast$ and $r_u^\ast$ significantly increase as $P_s$ increases. However, it is noteworthy that there is no solution within the prescribed range of $P_s$ according to Case 3. Since then, as $m$ and $m_s$ decrease from $2.5$ to $1.5$, the modified-$\mathcal{F}$ fading channel achieves a lower optimal value of $h^\ast$, resulting in a smaller maximum radius $r_u^\ast$ compared to moderate multipath fading and shadowing. Similarly, the maximum coverage area of the modified-$\mathcal{F}$ fading channel surpasses that of the conventional $\mathcal{F}$ fading channel no matter the channel experiences moderate fading or harsh fading. However, the optimal value of $r_u^\ast$ corresponding to the conventional $\mathcal{F}$ fading channel also increases when $m_s=1.5$, which is fallacious for channel conditions with harsh shadowing. 

Figure~\ref{Fig.hopt_zu} compares the flight altitude $h^\ast$ and service radius $r_u^\ast$ of the modified-$\mathcal{F}$ fading channel and the conventional $\mathcal{F}$ fading channel with respect to $z_c$ under different fading parameters. 
It can be observed that the curves of optimal value of $h^\ast$ regarding modified-$\mathcal{F}$ fading channel under different fading parameters all remain constant at 100m when the RIS-equipped-UAV is fairly close to the command center or the UE, but otherwise, they increase first and then decrease with the increase of $z_c$. Correspondingly, the optimal value of $r_u^\ast$ obtained from Case 2 severely decreases as $z_c$ increases. The trend of the curves obtained from Case 1 is consistent with $h^\ast$, which gradually increases first and then decreases. Likewise, there is no solution given by Case 3. 
As the channel experiences more harsh multipath fading and shadowing, the optimal values of $r_u^\ast$ and $h^\ast$ become smaller compared with those of moderate multipath fading and shadowing. Also, the optimal value of $r_u^\ast$ corresponding to the conventional $\mathcal{F}$ fading channel is smaller than that of the modified-$\mathcal{F}$ fading channel with $m_s=2.5$, while it increases as the shadowing becomes more harsh. 
In conclusion, it is crucial to select the optimal values of $h^\ast$ and $\theta^\ast$ according to various channel conditions, average transmit power, and distance for obtaining the maximum coverage area. 

\section{Conclusions}\label{sec:Conclusion}
In this paper, 
to accurately describe the channel under the effect of the topographical changes caused by secondary disasters and debris, a novel modified-Fisher-Snedecor $\mathcal{F}$ fading channel model and path-loss model were developed. Hereafter, we derived closed-form expressions for the newly revealed channel statistics in the cases of small and large number of reflecting elements, respectively. Furthermore, we presented analytical expressions for the average capacity, energy efficiency, and outage probability. Based on the above results, two optimization problems were formulated and solved. Specifically, we optimized the system energy efficiency with the constraint of RIS element number and optimized the coverage area subject to the altitude of RIS-equipped UAV. 
Simulation results showed that under our novel modified-Fisher-Snedecor $\mathcal{F}$ fading channel, the optimal number of RIS elements scheme and altitude of RIS-equipped-UAV scheme can achieve higher gain than the solutions to conventional Fisher-Snedecor $\mathcal{F}$ fading channel. 

\begin{appendix}
Based on the exact distribution of ${h_{_{u,i}}}$ and according to \cite[Eq.~(9.121.1)]{2007Table}, the PDF of the channel coefficient ${h_{_{u,i}}}$ given in Eq.~(\ref{hPDF}) can be rewritten as follows:
	\begin{equation}\label{rewrite}
	\begin{aligned}
	f({h_{_{u,i}}})&=\frac{{2\left(\frac{m\Omega_s}{m_s\Omega_m}\right)^{m} {h_{_{u,i}}^{2m-1}}}}{B(m,m_s)}\\
&\times_2F_1\left(m+m_s,m;m;-\frac{m\Omega_s{h_{_{u,i}}^2}}{m_s\Omega_m}\right),
	\end{aligned}
	\end{equation}
	where $_2F_1(\alpha, \beta; \gamma; -z)$ is the hypergeometric function \cite [Eq.(9.111)]{2007Table}. When ${h_{_{u,i}}}\rightarrow 0$, Eq.~(\ref{rewrite}) can be upper-bounded as follows: 
\begin{equation}\label{asy_h}
f({h_{_{u,i}}})\le\frac{2\left({\frac{m\Omega_s}{m_s\Omega_m}}\right)^{m}{h_{_{u,i}}^{2m-1}}}{B(m,m_s)}.
\end{equation}
Then, the Laplace transform of Eq.~(\ref{asy_h}) can be obtained as follows\cite{2007Table}:

\begin{equation}
\mathcal{L}_{h_{_{u,i}}}(s)=\frac{2\left({\frac{m\Omega_s}{m_s\Omega_m}}\right)^{m}\Gamma(2m)}{B(m,m_s)s^{2m}}.
\end{equation}
For the case of i.i.d Fisher-Snedecor $\mathcal{F}$ RVs, the Laplace transform of $A$ can be obtained as follows:
\begin{equation}
\mathcal{L}_{A}(s)=\left[\frac{2\left({\frac{m\Omega_s}{m_s\Omega_m}}\right)^{m}\Gamma(2m)}{B(m,m_s)s^{2m}}\right]^N.
\end{equation}
Thus, the upper bound of PDF for $A$, denoted by $f_{ub}(a)$, can be derived based on the Laplace inverse transform of $\mathcal{L}_{A}(s)$ as follows: 
\begin{equation}
f_{ub}(a)=\mathcal{L}_{A}^{-1}(s)=\left[\frac{2\left({\frac{m\Omega_s}{m_s\Omega_m}}\right)^{m}\Gamma(2m)}{B(m,m_s)}\right]^N\frac{a^{2Nm-1}}{\Gamma(2Nm)},
\end{equation}
and the upper bound of CDF for $A$, denoted by $F_{ub}(a)$, is expressed as follows:
\begin{equation}
F_{ub}(a)=\left[\frac{2\left({\frac{m\Omega_s}{m_s\Omega_m}}\right)^{m}\Gamma(2m)}{B(m,m_s)}\right]^N\frac{a^{2Nm}}{2Nm\Gamma(2Nm)}.
\end{equation}
 Since $a=\sqrt{\frac{\gamma {PL_{\tiny{C}}PL_{\tiny{U}}}}{{\gamma_0}}}$ holds, the diversity order is equal to $Nm$. 
\end{appendix}
\bibliographystyle{IEEEtran}
\bibliography{Ref}
\end{document}